\documentclass[12pt]{article}

\usepackage{jheppub}
\usepackage{graphicx}
\usepackage{amsmath,amssymb,amsfonts}
\usepackage{mathrsfs}
\usepackage{latexsym}
\usepackage{bm}
\usepackage{bbm}

\usepackage{subfigure}

\newcommand{\TeV}{\mbox{ TeV}}
\newcommand{\GeV}{\mbox{ GeV}}

\title{\Large\bf Strongly first order phase transition in the singlet fermionic dark matter model after LUX}

\author{Tai Li}
\author{and Yu-Feng Zhou}
\affiliation{ \textit{State Key Laboratory of Theoretical Physics,} \\
     \textit{Kavli Institute for Theoretical Physics China,} \\
     \textit{Institute of  Theoretical Physics, Chinese Academy of Sciences,} \\
     \textit{Beijing, 100190, P.~R.~China}}

\emailAdd{litai@itp.ac.cn}
\emailAdd{yfzhou@itp.ac.cn}

\abstract{
We investigate an extension of the standard model (SM) with a singlet fermionic dark matter (DM) particle which interacts with the SM sector through a real singlet scalar. The presence of a new scalar provides the possibility of generating a strongly first order phase transition needed for electroweak baryogenesis. Taking into account the latest Higgs search results at the LHC and the upper limits from the DM direct detection experiments especially that from the LUX experiment, and combining the constraints from the LEP experiment and the electroweak precision test,
we explore the parameter space of this model which can lead to the strongly first order phase transition.
Both the  tree- and loop-level barriers are included in the calculations.
We find that the allowed mass of the second Higgs particle is in the range $\sim 30-350\GeV$. The allowed mixing angle $\alpha$ between the SM-like Higgs particle and the second Higgs particle is constrained to $\alpha \lesssim 28^{\circ}$. The DM particle mass is predicted to be in the range $\sim 15-350\GeV$. The future XENON1T experiment can rule out a significant proportion of the parameter space of this model. The constraint can be relaxed only when the mass of the SM-like Higgs particle is degenerate with that of the second Higgs particle, or the mixing angle is small enough.
}

\begin{document}

\maketitle

\flushbottom

\section{Introduction}

The possibility of baryogenesis through electroweak phase transition (EWPhT) has been studied extensively (for reviews see e.g. Refs.~\cite{Cohen:1993nk,Rubakov:1996vz,Trodden:1998ym,Quiros:1999jp}). If the EWPhT is strongly first order, it can fulfill the condition of departure from thermal equilibrium which is one of the three conditions necessary for the generation of baryon number asymmetry in the Universe \cite{Sakharov:1967dj,Kuzmin:1985mm}. In order to avoid the washout of the baryon number asymmetry, the baryon number violating interactions induced by electroweak sphalerons must be suppressed at the temperature when the bubbles enveloping the broken phase start to nucleate~\cite{DeSimone:2011ek}. A commonly adopted assumption is that the sphaleronic interactions are suppressed immediately after the EWPhT, which leads to a requirement that $\varphi_c$ the vacuum expectation value (VEV) of the Higgs field in the broken phase is larger than the critical temperature, namely~\cite{Shaposhnikov:1987tw,Shaposhnikov:1986jp}
\begin{equation} \label{eq1}
\frac{\varphi_c}{T_c} \gtrsim \mathcal{E}.
\end{equation}
where $\mathcal{E}\approx 1$ is a constant. In the standard model (SM), the condition in Eq.~(\ref{eq1}) is satisfied only when the Higgs boson is very light, i.e., $m_h \lesssim 30 \GeV$ for $\mathcal{E} = 1$ \cite{Carrington:1991hz,Anderson:1991zb,Arnold:1992fb,Arnold:1992rz,Dine:1992wr}, which is ruled out by the current experiments, especially after the discovery of a $125\GeV$ Higgs boson at the LHC \cite{Chatrchyan:2012ufa,Aad:2012tfa}.
Thus new physics beyond the SM must be introduced for a successful electroweak baryogenesis.

Another clear indication of new physics is the existence of dark matter (DM),
which has been well established by astrophysical and cosmological observations as well as N-body simulations. According to the latest analysis reported by the Planck Collaboration, the measured energy density of DM in the Universe is
\cite{Ade:2013zuv}
\begin{equation} \label{omegah2}
\Omega h^2 = 0.1187\pm0.0017.
\end{equation}
Although the SM has been very successful in phenomenology, it can provide neither a strongly first order EWPhT
for baryogenesis nor a valid candidate of  DM.

One of the simplest models with DM candidates is the extension of the SM with a gauge singlet scalar field
\cite{Silveira:1985rk,McDonald:1993ex,Burgess:2000yq,Davoudiasl:2004be,He:2009yd,Gonderinger:2009jp,Bandyopadhyay:2010cc,Guo:2010hq,Mambrini:2011ik}.
The stability of the scalar can be protected by an $ad\ hoc$ $Z_2$ symmetry.
The $Z_2$ symmetry may be a residual symmetry from a global or local $U(1)$.
In the extension of the left-right symmetric models with a gauge singlet scalar, the $Z_2$ symmetry may originate from  the parity and CP symmetries \cite{Guo:2008si,Guo:2010vy,Guo:2010sy,Guo:2011zze,Liu:2012bm}.
However, if EWPhT is also required, it was shown that the singlet scalar could contribute only up to $3\%$ of the DM energy density \cite{Cline:2012hg,Cline:2013gha}.
In the inert doublet model, an additional $SU(2)$ doublet is added to the SM.
This model can provide a valid DM candidate
and also trigger strongly first order EWPhT,
	due to the contributions from other charged and neutral scalars in the additional doublet \cite{Chowdhury:2011ga}.
When taking into account  the data of the LHC and DM direct detection experiments,
the parameter space of this model is highly constrained \cite{Borah:2012pu,Arhrib:2013ela}.

The DM particle can also be  a gauge singlet fermion
which interacts with the SM sector through a gauge singlet real scalar.
The phenomenology of this type of DM model has been explored in Refs. \cite{Kim:2008pp,Lee:2008xy,Kim:2009ke,Baek:2011aa,Baek:2012uj}.
Light subGeV-scale singlet scalars exchanged by the fermionic DM particles
can serve as a force-carrier  in the mechanism of
the Sommerfeld enhancement which has been considered to explain the large boost
factors suggested  by the data of various DM indirect detection experiments
(see e.g. Refs.
\cite{
Sommerfeld31,Hisano:2002fk,Hisano:2003ec,Cirelli:2007xd,ArkaniHamed:2008qn,
Pospelov:2008jd,MarchRussell:2008tu,Cassel:2009wt,
Liu:2013vha,Chen:2013bi}), such as
PAMELA \cite{Adriani:2008zr},
Fermi-LAT \cite{Abdo:2009zk,1109.0521} and AMS-02 \cite{PhysRevLett.110.141102}
(for a recent analysis see e.g. \cite{Jin:2013nta}).

It is of interest to investigate whether the strongly first order EWPhT can also be realized in
	the singlet fermionic DM model.
This question was addressed in Ref \cite{Fairbairn:2013uta} in which the discussion
	was limited to the case of tree-level barrier only. However, without the $Z_2$ symmetry, the strongly first order EWPhT can be achieved from the singlet scalar contributions via both tree- and loop-level effects due to the linear and cubic terms in the singlet scalar and Higgs potential, which is similar with the case of the SM plus a gauge singlet real scalar \cite{Espinosa:2011ax,Chung:2012vg,Choi:1993cv,Ham:2004cf,Ahriche:2007jp,Profumo:2007wc,Cline:2009sn,Huang:2012wn}.
In this work, we aim at an extensive and up-to-date analysis of the EWPhT in this model.
In comparison with the previous analysis, we make the following improvements:
\begin{itemize}
  \item
We go beyond the tree-level analysis by  including  the loop-level barrier induced from the
thermal corrections to  the effective potential.
We show that when taking into account both the tree- and loop-level barriers the allowed parameter space  is
significantly enlarged.
For instance, the upper limit on the mass of the second Higgs particle is about $100\GeV$ higher at $\sin\alpha=0.001$.
At the same time the critical temperature after including the cubic terms from one-loop corrections
is about $10\%$ higher.
We show that in this case the allowed  mass of the second Higgs particle can reach $\sim 600\GeV$.

 \item We adopt an improved analytical approximation of the finite temperature effective potential which well matches both the usual high- and low-temperature approximations. This approximation makes our analysis valid for large values of $\varphi_c/T_c$, which is of crucial importance as the value of $\varphi_c/T_c$ can reach up to $10$ in this model.

 \item We consider the contribution from the sphaleron magnetic moment to the sphaleron energy. We find that in this model the contribution from the sphaleron magnetic moment is weakened compared with the case of the SM, due to the extra scalar field. The sphaleron magnetic moment energy can lead to a difference between the values of $\varphi_c/T_c$ and $E_{\text{sph}}(T_c)/35T_c$ within $10\%$.

  \item We include the latest upper limits on DM-neucleon scattering cross section from the LUX experiment \cite{Akerib:2013tjd} which is about one order of magnitude stronger than the previous one reported by XENON100 \cite{Aprile:2012nq}. As a consequence the mixing angle between the SM-like and the second Higgs  particles
 is stringently constrained.

  \item We focus on the constraints on the phenomenologically interesting physical parameters such as  the mass of the second Higgs particle, the mixing angle and the DM particle mass.
A numerical scan of the parameter space of this model is performed using a Markov Chain Monte Carlo (MCMC) approach.
Taking into account the latest data from  the LHC and the LUX experiments, and combining the constrains from the LEP experiment and the electroweak precision test, we find that the mass of the second Higgs particle is in the range $\sim 30 - 350\GeV$ and the mixing angle is constrained to $\alpha \lesssim 28^{\circ}$. We also find that the DM particle mass is predicted to be in the range $\sim 15-350\GeV$.
\end{itemize}

This paper is organized as follows. We first give a brief overview of the singlet fermionic DM model in section~\ref{smodel}. In section~\ref{sewpt}, we discuss the effective potential at finite temperature at the tree- and loop-level. A numerical analysis of parameter space is performed and the allowed parameter space is given in section~\ref{spara}. In section~\ref{sec:sphaleron} we discuss the correction of the sphaleron energy from the magnetic dipole and its effect on the parameter space allowed by the requirement of a strongly enough first order EWPhT. We then investigate the constraints from DM thermal relic density (section~\ref{srelic}), DM direct detection (section~\ref{sdire}), LHC data on Higgs signal strength (section~\ref{sdiph}), LEP data and electroweak precision test (section~\ref{slep}). The combined result is present in section~\ref{sresu}. Finally, conclusions and some discussions are given in section~\ref{sconc}.

\section{Singlet fermionic dark matter model} \label{smodel}

We consider an extension of the SM with a gauge singlet Dirac fermion $\psi$ which interacts with SM particles through a gauge singlet scalar $S$.
The tree-level Higgs potential of this model is given by
\begin{equation} \label{eq:Ls}
V\left(\Phi,S\right)= -\mu_\phi^2 \Phi^\dagger \Phi + \lambda_\phi \left(\Phi^\dagger\Phi\right)^2 - \mu_1^3 S - \frac{1}{2}\mu_{s}^{2}S^{2} - \frac{1}{3}\mu_{3}S^{3} + \frac{1}{4}\lambda_{s}S^{4}+\mu \Phi^{\dagger}\Phi\,S + \frac{1}{2} \lambda \Phi^{\dagger}\Phi\,S^2,
\end{equation}
where $\Phi$ is the SM Higgs doublet
\begin{equation}
\Phi=\left(\begin{array}{c}
G^{+}\\
\frac{1}{\sqrt{2}}\left(\phi^0-iG^0\right)
\end{array}\right),
\end{equation}
where $G^\pm$, $G^0$ are the would-be Goldstone bosons. The coefficient $\mu_1$ in Eq.~(\ref{eq:Ls}) can be eleminated by a shift of the field $S$, $S\rightarrow S+\sigma$, which only causes a redefinition of parameters. In general both $\phi^0$ and $S$ can develop non-zero VEVs at zero temperature which are defined as $\varphi_0 \equiv \langle \phi^0 \rangle\mid_{T=0}$ and $s_0 \equiv \langle S \rangle\mid_{T=0}$. The last two terms in Eq.~(\ref{eq:Ls}) lead to off-diagonal terms in the squared mass matrix of singlet scalar and the SM Higgs boson, which introduces a mixing between $\phi^0$ and $S$. The squared mass matrix of $\phi^0$ and $S$ is given by
\begin{equation} \label{massmatrix}
\mathcal{M}^{2} = \left(\begin{array}{cc}
\mathcal{M}_{11}^2 & \mathcal{M}_{12}^2 \\
\mathcal{M}_{21}^2 & \mathcal{M}_{22}^2
\end{array}\right),
\end{equation}
where
\begin{eqnarray} \label{mij}
\mathcal{M}_{11}^2 &=& -\mu_{\phi}^{2}+3\lambda_{\phi}\varphi_0^{2}+ \frac{1}{2} \lambda s_0^2 + \mu s_0 , \notag \\
\mathcal{M}_{22}^2 &=& -\mu_{s}^{2}-2\mu_{3}s_0+3\lambda_{s}s_0^{2}+ \frac{1}{2} \lambda \varphi_0^{2} , \\
\mathcal{M}_{12}^2 &=& \mathcal{M}_{21}^2 = \mu \varphi_0+\lambda \varphi_0s_0 . \notag
\end{eqnarray}
The squared mass matrix in Eq.~(\ref{massmatrix}) can be diagonalized by rotating $\phi^0$ and $S$ into mass eigenstates ($h$, $H$)
\begin{equation}
\left(\begin{array}{c}
h\\
H
\end{array}\right)=\left(\begin{array}{cc}
\cos\alpha & -\sin\alpha\\
\sin\alpha & \cos\alpha
\end{array}\right)\left(\begin{array}{c}
\phi^0\\
S
\end{array}\right),
\end{equation}
where the mixing angle $\alpha$ is
\begin{equation}  \label{tan2a}
\tan2\alpha = \frac{2 m_{12}^2}{\left(m_{22}^2-m_{11}^2\right)}.
\end{equation}
The value of $\alpha$ is defined in the range $0^\circ-45^\circ$, such that $h$ plays the role of the SM-like Higgs particle while $H$ is singlet dominant.
The interaction involving the singlet fermionic DM particle $\psi$ is given by the Lagrangian
\begin{equation}
\mathcal{L}_{\psi}=i\,\bar{\psi} \partial\!\!\!/ \psi - y_{\psi}\bar{\psi}\psi S.
\end{equation}
In general $S$ can develope a non-zero VEV, which contributes to the mass of the fermionic DM particle $\psi$. In this work we consider the case where $\psi$ only obtains mass from the VEV of $S$, namely $m_\psi=y_\psi s_0$, which makes the model more predictive.

\section{Effective potential and EWPhT} \label{sewpt}

The tree-level potential for $\varphi = \langle\phi^0\rangle$ and $s = \langle S\rangle$ can be written as
\begin{equation} \label{v0}
V_0\left(\varphi,s\right) = -\frac{1}{2} \mu_{\phi}^2 \varphi^2 - \frac{1}{2} \mu_s^2 s^2 - \frac{1}{3} \mu_3 s^3 + \frac{1}{2} \mu \, s \,\varphi^2 + \frac{1}{4} \lambda_{\phi} \varphi^4 + \frac{1}{4} \lambda_s s^4 + \frac{1}{4}\lambda \, s^2 \varphi^2 .
\end{equation}
The coefficients $\mu_{\phi}$ and $\mu_s$ can be rewritten in terms of the VEVs $\varphi_0$ and $s_0$ according to the minimization conditions of the tree-level potential. However, the minimization conditions can not guarantee that $\left(\varphi_0, s_0\right)$ is the global minimum. Thus a check on whether there exists a deeper minimum is needed. In order to guarantee the stability of $\left(\varphi_0, s_0\right)$ as the global vacuum, it is also required that the potential is bounded-from-below.

The parameters $\lambda_{\phi}$, $\mu$ and $\lambda$ can be rewritten in terms of three physical parameters, i.e. the masses of the two Higgs particles $m_h$, $m_H$ and the mixing angle $\alpha$, as follows
\begin{eqnarray}
\lambda_{\phi} & =& \frac{1}{2\varphi_0^{2}}\left(m_{h}^{2}\cos^{2}\alpha+m_{H}^{2}\sin^{2}\alpha\right) , \notag \\
\mu & =&-2\frac{s_0}{\varphi_0^{2}} \left(m_{h}^{2}\sin^{2}\alpha+m_{H}^{2}\cos^{2}\alpha+\mu_{3}s_0-2\lambda_{s}s_0^{2}\right) , \\
\lambda & =& \frac{1}{\varphi_0s_0}\left[\left(m_{H}^{2}-m_{h}^{2}\right)\sin\alpha\cos\alpha-\mu \varphi_0\right]. \notag
\end{eqnarray}

We include one-loop Coleman-Weinberg correction of the potential at zero temperature \cite{Coleman:1973jx}
\begin{equation} \label{v1}
V_1\left(\varphi,s\right) = \frac{1}{64\pi^2} \sum_{i} N_{i} m_{i}^4 \left(\varphi,s\right) \left[\log \frac{m_{i}^2 \left(\varphi,s\right)}{Q^2} - C_{i}\right],
\end{equation}
where $i$ runs over all the particles in the loop, and $N_{i}$ is the degrees of freedom of the particle $i$, $C_i$ is a constant ($C_i = 6/5$ for gauge bosons, $C_i = 3/2$ for scalars and fermions), $Q$ is the renormalization scale which we fix at the mass of the top quark. The counter terms $V_{\text{CT}}\left(\varphi,s\right)$ needed to renormalize the potential are given in Appendix~\ref{appA}.

The one-loop effective potential at finite temperature $T$ can be written as
\begin{equation} \label{veff}
V_{\text{eff}}\left(\varphi,s;T\right)=V_{0}\left(\varphi,s\right) + V_{1}\left(\varphi,s\right) + V_{\text{CT}}\left(\varphi,s\right) + V_{1}\left(\varphi,s;T \right) ,
\end{equation}
where $V_1\left(\varphi,s;T\right)$ is the one-loop thermal corrections
\begin{equation}
V_{1}\left(\varphi,s;T\right) = \frac{T^4}{2\pi^2} \left[ \underset{i}{\sum} n_i I_{\text{B}}\left(a_i\right) + \underset{j}{\sum} n_j I_{\text{F}}\left(a_j\right) \right],
\end{equation}
where $a=m^2\left(\varphi,s\right)/T^2$, $i$ ($j$) runs over all the bosons (fermions), $n_{i(j)}$ denotes the degrees of freedom of bosons (fermions), and $I_{{\text{B}}({\text{F}})} \left(a\right)$ is defined as
\begin{equation} \label{Ibf}
I_{{\text{B}}({\text{F}})}\left(a\right) = \int_0^{\infty} dx \, x^2 \ln \left(1 \mp e^{-\sqrt{x^2 + a}} \right),
\end{equation}
where the sign $-$ ($+$) is for bosons (fermions).

Since the evaluation of the integration in Eq.~(\ref{Ibf}) is computationally expensive, it is necessary to have an analytical approximation. In the high temperature limit, i.e. $m\left(\varphi,s\right)/T \ll 1$, $I_{{\text{B}}({\text{F}})}\left(a\right)$ can be expanded as \cite{Dolan:1973qd}
\begin{eqnarray} \label{IhighT}
I_{{\text{B}}}^{(1)} \left(a\right) &=& -\frac{\pi^4}{45} + \frac{\pi^2}{12} a - \frac{\pi}{6} a^{\frac{3}{2}} - \frac{1}{32} a^2 \left[\log\left(a\right) - \gamma_{\text{B}}\right] ,  \label{highTb} \\
I_{{\text{F}}}^{(1)} \left(a\right) &=& -\frac{7\pi^4}{360} + \frac{\pi^2}{24} a + \frac{1}{32} a^2 \left[\log\left(a\right) - \gamma_{\text{F}}\right] ,  \label{highTf}
\end{eqnarray}
where $\gamma_{\text{B}} = 5.40762$ and $\gamma_{\text{F}} = 2.63503$. The term cubic in $m/T$ in Eq.~(\ref{highTb}) gives rise to the barrier in the potential which makes the phase transition first order. In the low temperature limit, $I_{{\text{B}}({\text{F}})}\left(a\right)$ can be expanded as \cite{Anderson:1991zb}
\begin{equation}  \label{lowT}
I_{\text{B}}^{(2)}\left(a;n\right) = I_{\text{F}}^{(2)}\left(a;n\right) = - \sqrt{\frac{\pi}{2}} \, a^{\frac{3}{4}} \, e^{-a^{1/2}} \left( 1+\frac{15}{8}a^{\frac{1}{2}}+\frac{105}{128}a \right).
\end{equation}

\begin{figure}[!tb]
  \centering
  \subfigure{
    \label{fI:b} 
    \includegraphics[width=200pt]{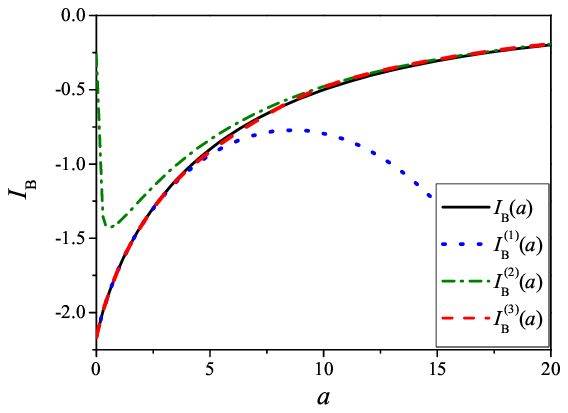}}
  \subfigure{
    \label{fI:f} 
    \includegraphics[width=200pt]{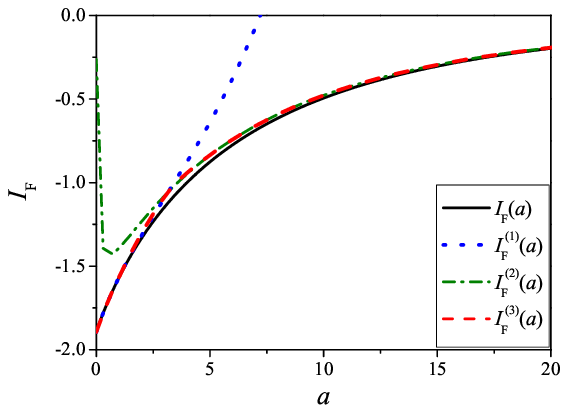}}
  \caption{Left) Comparison of different approximations of the function $I_{\text{B}}(a)$. The black solid and red dashed curves correspond to the numerical value and the approximation $I^{(3)}_{\text{B}}(a)$, respectively. The blue dotted and green dot-dashed curves correspond to the high- and low-temperature approximations, respectively. Right) The same as left but for $I_{\text{F}}(a)$.}
  \label{fI} 
\end{figure}

The high- and low-temperature approximations are shown in Fig.~\ref{fI}. It can be seen that the high temperature approximation starts to fail when $a \gtrsim 3$. By matching the high- and low-temperature approximations, we obtain a reasonable approximation to the integral
\begin{equation} \label{Iapp}
I_{{\text{B}}({\text{F}})}^{(3)}\left(a\right) = t_{{\text{B}}({\text{F}})}\left(a\right) \, I_{{\text{B}}({\text{F}})}^{(1)} \left(a\right) + \left(1-t_{{\text{B}}({\text{F}})}\left(a\right) \right) \, I_{{\text{B}}({\text{F}})}^{(2)} \left(a;2\right),
\end{equation}
where $t_{\text{B}}(a) = e^{-\left(a/6.3\right)^4}$ and $t_{\text{F}}(a) = e^{-\left(a/3.25\right)^4}$ are obtained by numerically fitting to the exact value of the integral. A comparison of different approximations of $I_{{\text{B}}({\text{F}})}(a)$ is shown in Fig.~\ref{fI}. For the approximation $I^{(3)}_{{\text{B}}({\text{F}})}(a)$ in Eq.~(\ref{Iapp}), the deviation to the exact value of $I_{{\text{B}}({\text{F}})}$ is less than $5\%$ in the region $0\leqslant a\leqslant 20$.

The calculation of effective potential can be further improved by including thermal corrections to the boson masses which come from high order ring diagrams. After including the ring diagrams, the field-dependent squared mass matrix for the two Higgs particles is given by
\begin{equation}
\mathcal{M}^2\left(\varphi, s; T\right) = \left(\begin{array}{cc}
\mathcal{M}_{11}^2 & \mathcal{M}_{12}^2 \\
\mathcal{M}_{21}^2 & \mathcal{M}_{22}^2
\end{array}\right) + \left( \begin{array}{cc}
c_\phi & 0 \\
0 & c_s
\end{array} \right) T^2,
\end{equation}
where the matrix elements $\mathcal{M}_{ij}$ are defined analogously as in Eq.~(\ref{mij}) with replacements $\varphi_0\rightarrow\varphi$, $s_0 \rightarrow s$, $c_\phi$ and $c_s$ are defined as
\begin{eqnarray}
c_\phi &=& \frac{1}{48}\left(9g^2 + 3g^{\prime 2} + 12 y_t^2 + 24 \lambda_{\phi} + 2 \lambda\right), \\
c_s &=& \frac{1}{12}\left(2\lambda + 3 \lambda_s + 2 y_{\psi}^2\right),
\end{eqnarray}
where $y_t$ is the top Yukawa coupling, $g$ and $g^{\prime}$ are the $SU(2)_L$ and $U(1)_Y$ gauge couplings, respectively. The thermal masses of the Goldstone bosons are given by
\begin{equation}
m_{G^0,G^{\pm}}^2\left(\varphi, s; T\right) = -\mu_{\phi}^2 + \lambda_{\phi} \varphi^2 + \mu s + \frac{1}{2} \lambda s^2 + c_\phi T^2.
\end{equation}

In order to trigger first order EWPhT, the thermal effective potential must have two degenerate minima separated by a barrier at the critical temperature. Due to the existence of the extra scalar field, there can exist two kinds of barriers in this model
\begin{itemize}
  \item {\bf Tree-level barrier.} This kind of barrier arises from the terms linear and cubic in $s$ which are already present in the effective potential at tree-level. In the scenario with tree-level barrier only, one important implication is that a first order EWPhT is always related to a change of the VEV of the singlet scalar field at the critical temperature. If the VEV of the singlet scalar field is constant during the EWPhT, the tree-level potential would have the same structure as that in the SM case which has no barrier.
  \item {\bf Loop-level barrier.} This kind of barrier arises from the term cubic in $m/T$ which comes from the thermal one-loop corrections of the bosonic fields to the effective potential. It also exists in the SM case, which is however not enough to trigger a strongly first order EWPhT. In this model, the extra singlet scalar field can contribute to this kind of barrier and make it possible to trigger a strongly first order EWPhT.
\end{itemize}

For the investigation of the tree-level barrier, it is enough to keep only the leading order terms which are quadratic in $m/T$ of the high-temperature approximation
\begin{equation} \label{V1lo}
V_1^{\text{lo}}\left(\varphi, s; T\right) = \left(\frac{1}{2} \kappa_\phi \varphi^2 + \frac{1}{2} \kappa_s s^2 + \kappa_3 s \right) T^2,
\end{equation}
where
\begin{eqnarray}
\kappa_\phi &=& \frac{1}{48} \left(9g^2 + 3g^{\prime 2} + 12 y_t^2 + 24 \lambda_{\phi} + 2\lambda \right), \notag \\
\kappa_s &=& \frac{1}{12} \left(2\lambda + 3 \lambda_s + 2 y_{\psi}^2 \right), \\
\kappa_3 &=& \frac{1}{12} \left(-\mu_3 + 2 \mu\right). \notag
\end{eqnarray}

For an illustration of the tree-level barrier, we use $V_0\left(\varphi,s\right)+ V_1^{\text{lo}}\left(\varphi,s;T \right)$ as an approximation of the effective potential. The stationary points of this effective potential are located at the intersections of the curves determined by $\partial V_{\text{eff}}\left(\varphi, s;T\right) / \partial \varphi = 0$ and $\partial V_{\text{eff}}\left(\varphi, s;T\right) /\partial s = 0$ which lead to
\begin{equation} \label{path1}
\varphi = 0 \quad \text{or} \quad \varphi^2 = f_h\left(s\right) = -\frac{\kappa_\phi T^2 - \mu_\phi^2 + \mu s + \frac{1}{2}\lambda s^2}{\lambda_\phi},
\end{equation}
and
\begin{equation} \label{path2}
\varphi^2 = f_s\left(s\right) = -2\cdot\frac{\kappa_3 T^2 + \left(\kappa_s T^2-\mu_s^2\right) s - \mu_3 s^2 +\lambda_s s^3}{\mu+\lambda s}.
\end{equation}

\begin{figure}[!bt]
  \centering
  \begin{minipage}[c]{.15\textwidth}
    \centering
    \includegraphics[width=50pt]{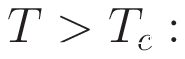}
  \end{minipage}
  \begin{minipage}[c]{.4\textwidth}
    \centering
  \subfigure[]{ \label{fhightem}
    \includegraphics[width=160pt]{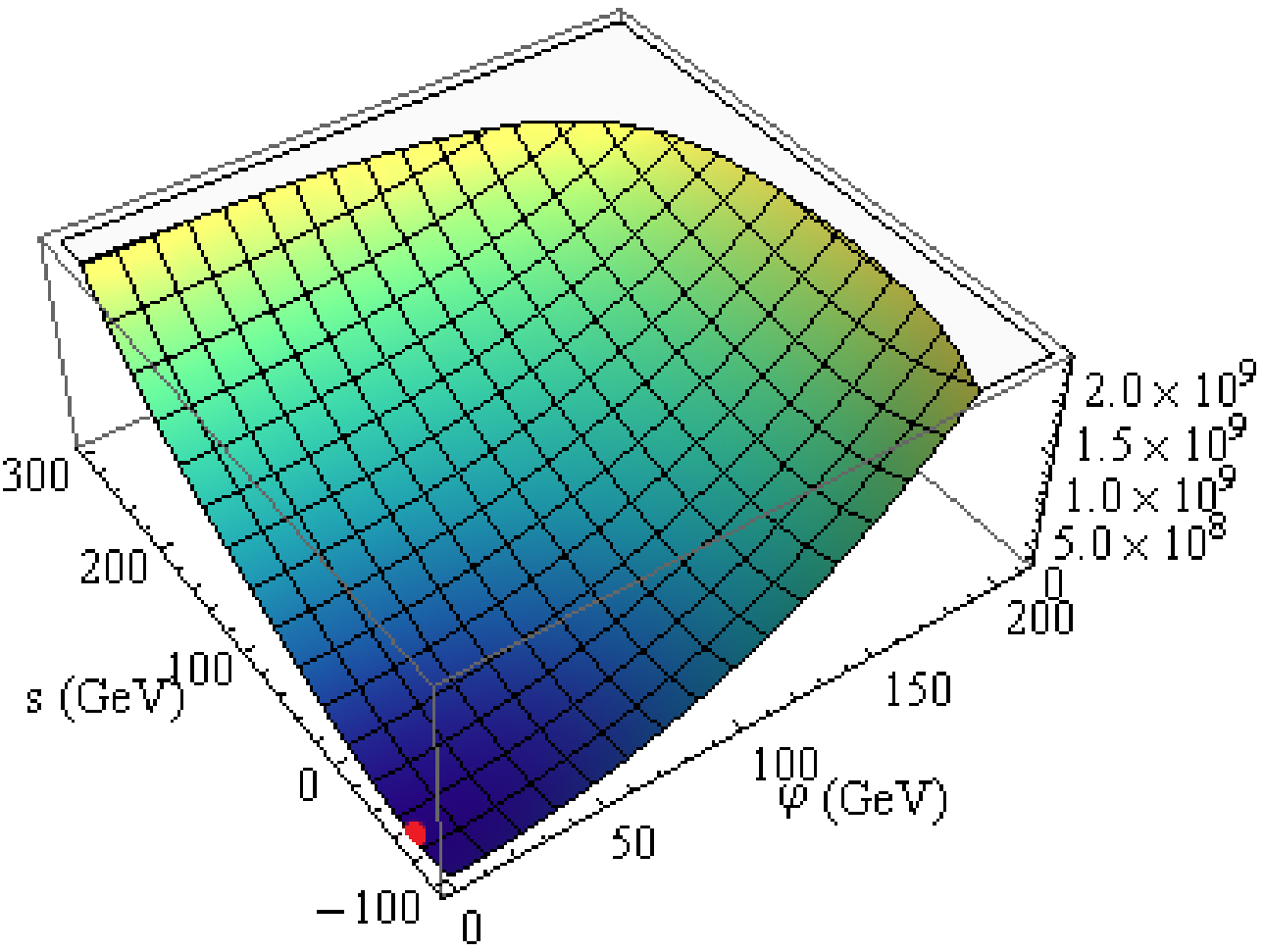}}
  \end{minipage}
  \begin{minipage}[c]{.4\textwidth}
    \centering
  \subfigure[]{
    \includegraphics[width=160pt]{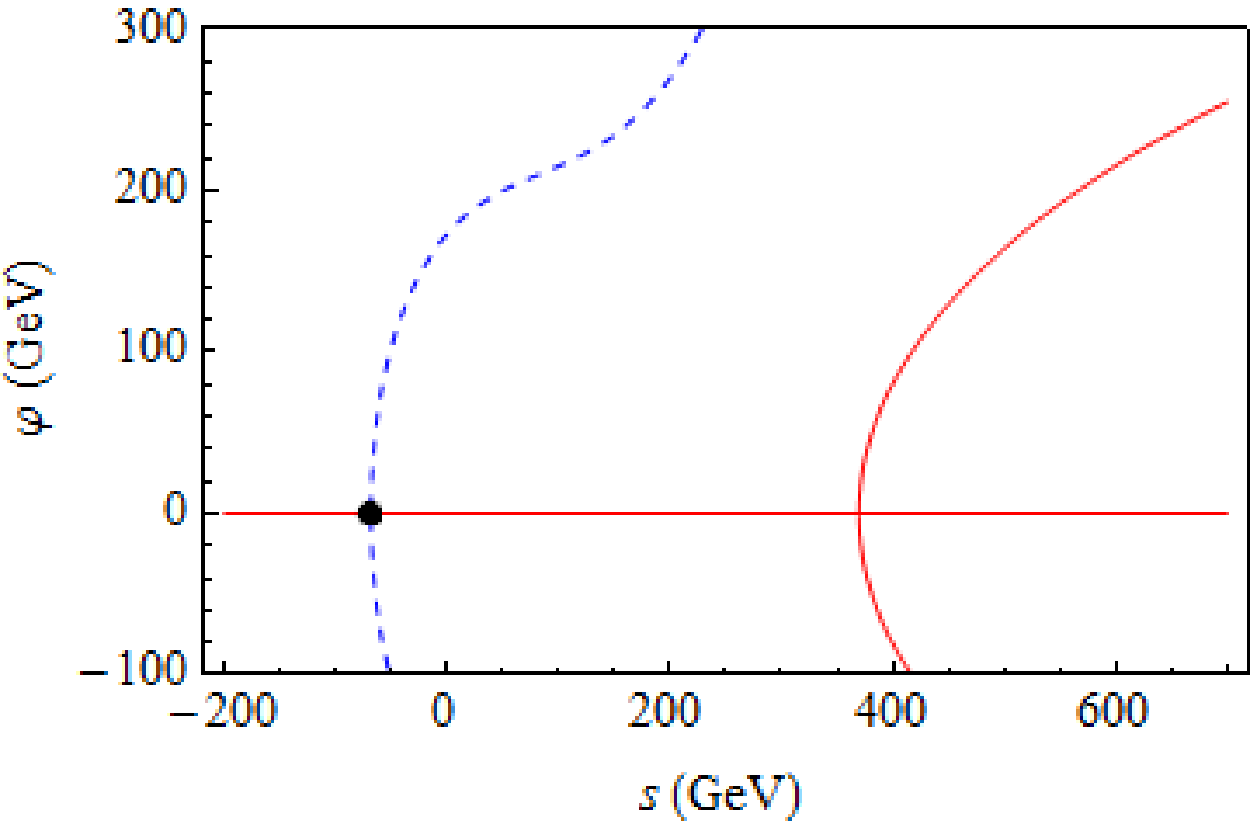}}
  \end{minipage}

  \begin{minipage}[c]{.15\textwidth}
    \centering
    \includegraphics[width=50pt]{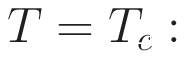}
  \end{minipage}
  \begin{minipage}[c]{.4\textwidth}
    \centering
  \subfigure[]{ \label{fcriticaltem}
    \includegraphics[width=160pt]{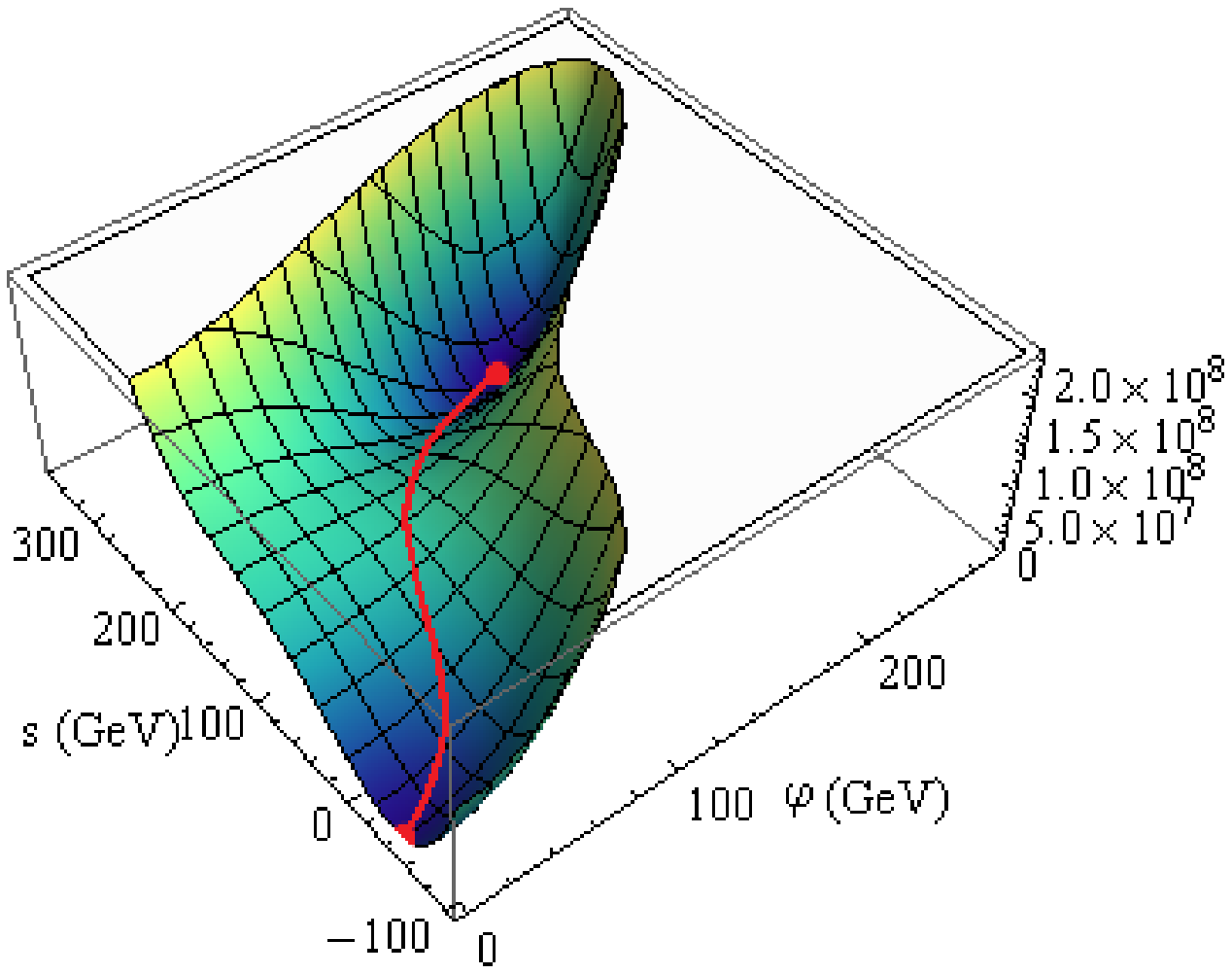}}
  \end{minipage}
  \begin{minipage}[c]{.4\textwidth}
    \centering
  \subfigure[]{
    \includegraphics[width=160pt]{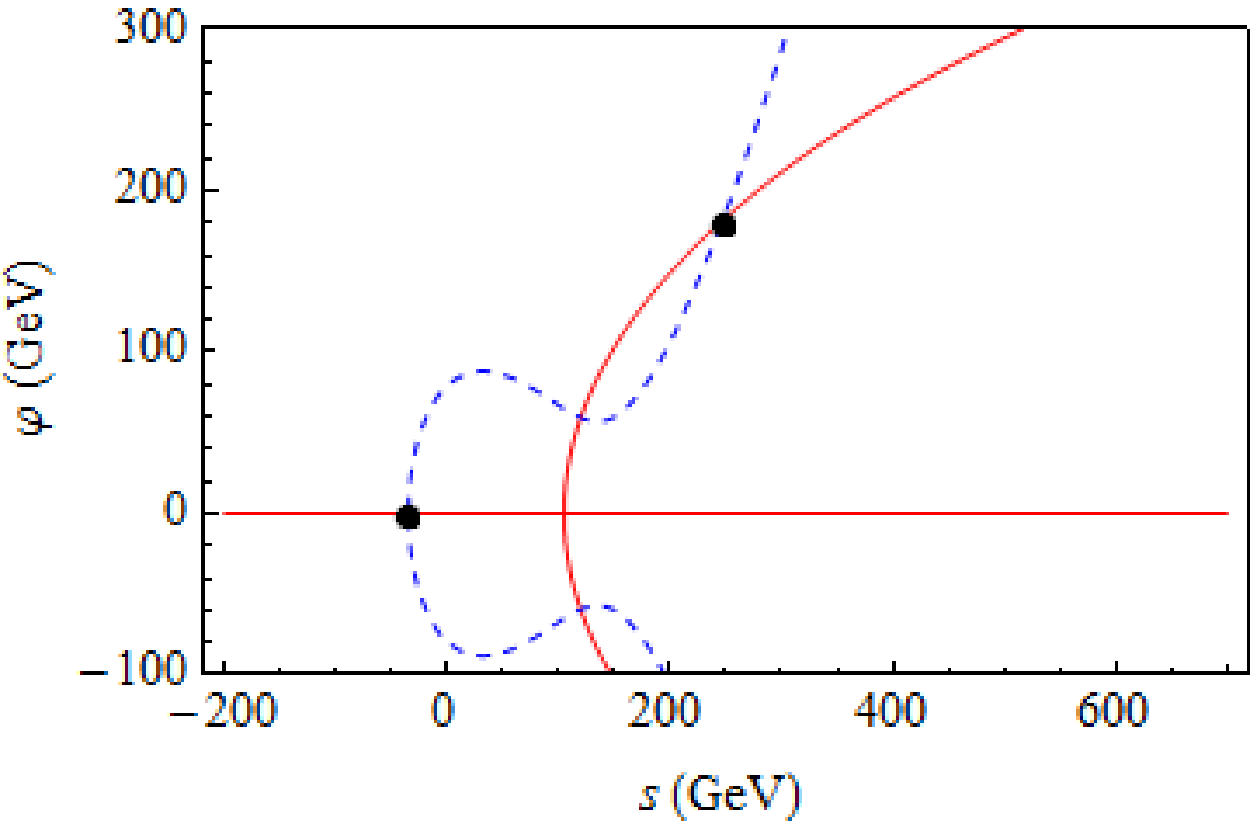}}
  \end{minipage}

  \begin{minipage}[c]{.15\textwidth}
    \centering
    \includegraphics[width=50pt]{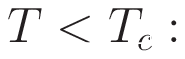}
  \end{minipage}
  \begin{minipage}[c]{.4\textwidth}
    \centering
  \subfigure[]{ \label{flowtem}
    \includegraphics[width=160pt]{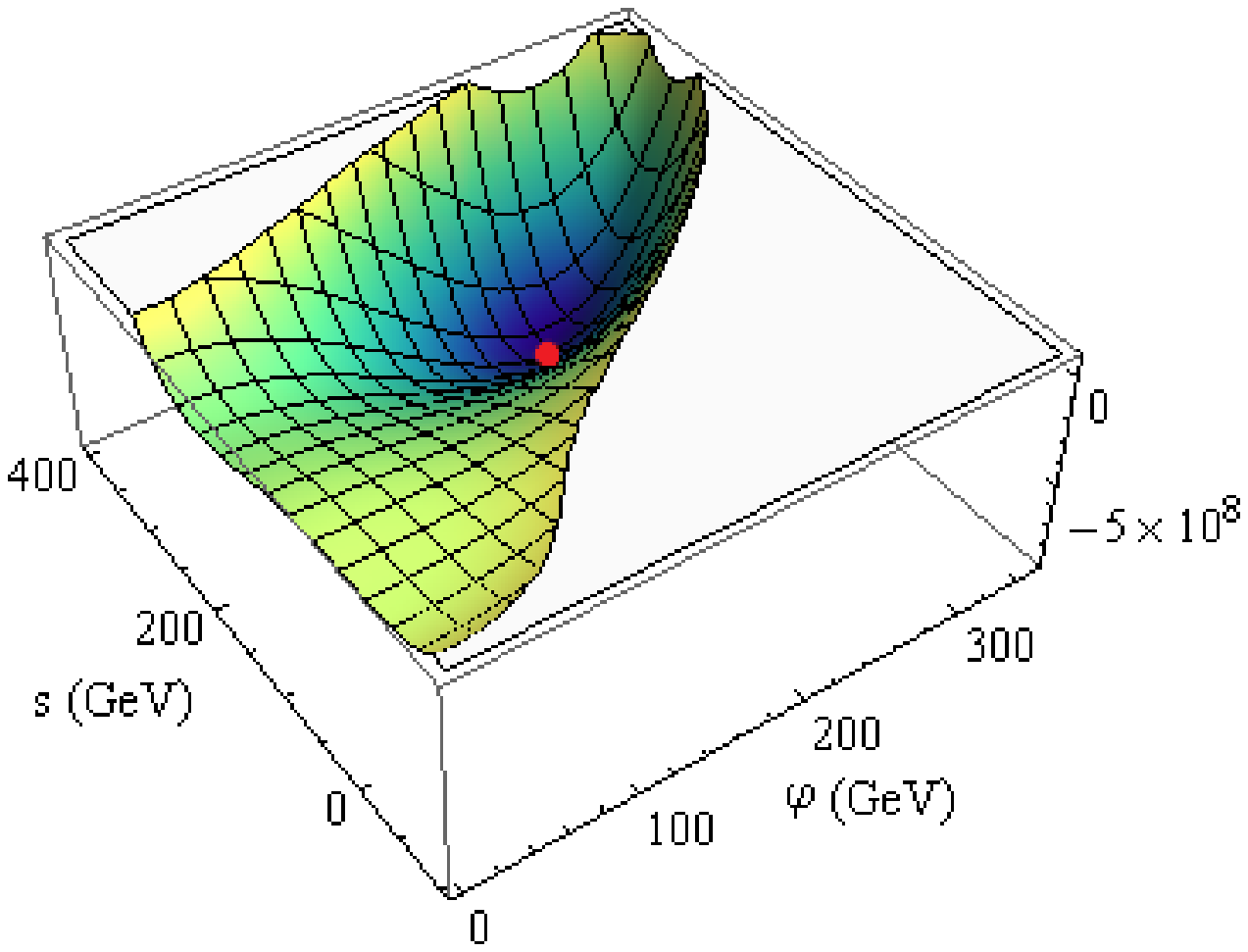}}
  \end{minipage}
  \begin{minipage}[c]{.4\textwidth}
    \centering
  \subfigure[]{
    \includegraphics[width=160pt]{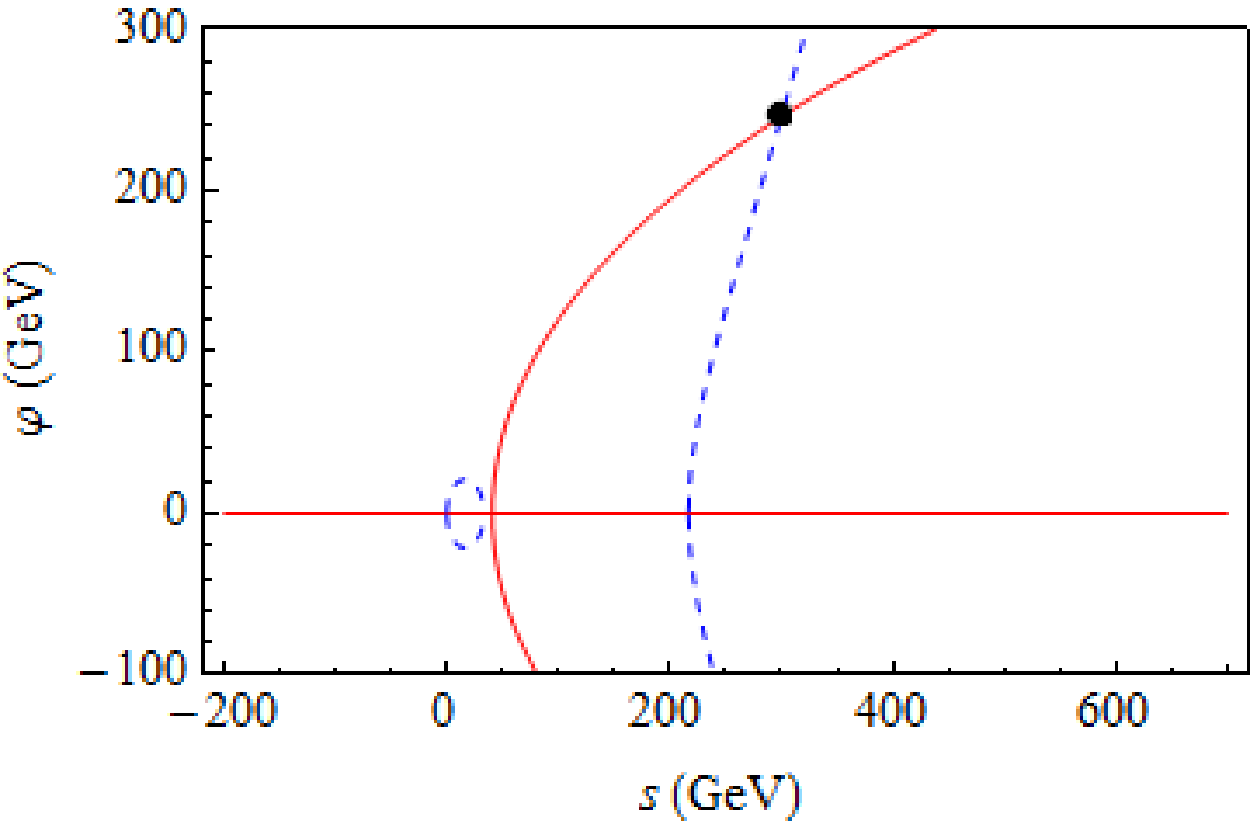}}
  \end{minipage}

  \caption{Thermal evolution of the effective potential in the senario with tree-level barrier only. The parameters are fixed at $s_0=300\GeV$, $\lambda_\phi = 1$, $\mu_3 = 250\GeV$, $\lambda_s = 1$, $\mu = -250\GeV$, $\lambda = 0.1$ and $y_{\psi} = 0.5$. Left) The effective potentials at $T>T_c$, $T=T_c$, and $T<T_c$ from top to bottom, respectively. The global minima of the effective potentials are indicated by red dots. In (c) the path with lowest barrier between the two local minima is indicated by the red line. Right) Curves corresponding to $\partial V_{\text{eff}}\left(\varphi, s;T\right) / \partial \varphi = 0$ (solid line) and $\partial V_{\text{eff}}\left(\varphi, s;T\right) /\partial s = 0$ (dashed line). The global minima are located at the intersections of the two curves as indicated by the black dots.}
  \label{fefftems} 
\end{figure}

We show the evolution of this effective potential with temperature in Fig.~\ref{fefftems}. Since at sufficiently high temperature the effective potential is dominated by the contributions from the thermal corrections in Eq.~(\ref{V1lo}), there is only one minimum at $\varphi = 0$, as shown in Fig.~\ref{fhightem}. As the temperature decreases, local minimum with $\varphi\neq0$ appears, but the original minimum at $\varphi = 0$ is still the global one. At the critical temperature $T_c$, the minimum at $\varphi = \varphi_c$ becomes degenerate with the minimum at $\varphi = 0$, as shown in Fig.~\ref{fcriticaltem}. The minimum at $\varphi = 0$ becomes meta-stable and the phase transition of $\varphi$ occurs. It can be seen that there is a barrier which separates the two degenerate minima and leads to first order EWPhT. After the phase transition of $\varphi$, the local minimum at $\varphi\neq0$ becomes the global one, as shown in Fig.~\ref{flowtem}.

\section{Parameter space for EWPhT} \label{spara}

To check whether a EWPhT is strongly first order, we should first find the critical temperature which is defined as when there appear two degenerate minima. We search for $T_c$ in the range from $T_{\text{min}}=1\GeV$ to $T_{\text{max}}=1\TeV$. We start from $T_{\text{min}}$, then increase the temperature and check the minima of the potential. The critical temperature is obtained when the local minimum at $\varphi\neq0$ becomes degenerate with the one at $\varphi=0$. If the global minimum at $T_{\text{max}}$ is at $\varphi\neq0$, EWPhT will not occur.

When the EWPhT occurs, there is a path connecting the two degenerate local minima which has the lowest barrier (see Fig.~\ref{fcriticaltem}). If there is no barrier along this path, the EWPhT is of the second order. In this case the local minimum corresponds to a flat direction of the potential. To identify this case we follow the method in Ref.~\cite{Cline:2009sn} to check whether a putative minimum is a real minimum. We minimize the potential on small circles surrounding the putative local minimum. If the minima on the circles are greater than the putative minimum, it is indeed a true local minimum.

We explore the full parameter space of the singlet fermionic DM model which includes: $m_H$, $\alpha$, $s_0$, $\mu_3$, $\lambda_s$, and $m_{\psi}$. We scan these parameters in the ranges
\begin{eqnarray}
10\GeV \leqslant m_H \leqslant 1\TeV, \quad & 0 \leqslant \alpha \leqslant 45^{\circ}, \quad & -1\TeV < s_0 \leqslant 1\TeV, \notag \\
-1\TeV \leqslant \mu_3 \leqslant 1\TeV, \quad & 0 \leqslant \lambda_s \leqslant 3, \quad & -3 \leqslant y_{\psi} \leqslant 3.
\end{eqnarray}
The mass of the SM-like Higgs particle is fixed at $m_h = 125\GeV$.

We use an improved random walk sampling algorithm to scan the parameter space based on a MCMC method with the Metropolis algorithm. The likelihood of a given parameter set $\boldsymbol{x}$ is defined as
\begin{equation} \label{lik}
\mathscr{L}\left(\boldsymbol{x}\right) = \min\{\varphi_c/T_c, 1\}.
\end{equation}
We run multi-chain samplers with initial values uniformly distributed in the 6-dimensional parameter space and obtain a sample set containing about $5\times10^6$ sample points satisfying $\varphi_c/T_c>1$.

\begin{figure}[!bt]
  \centering
  \includegraphics[width=200pt]{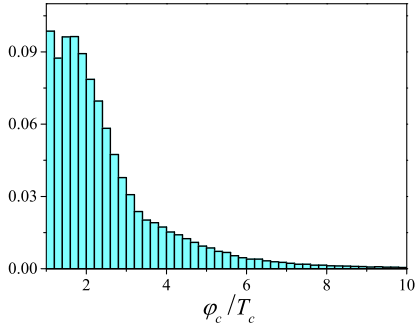}
  \caption{The relative frequency distribution of the order parameter $\varphi_c/T_c$ of the samples satisfying $\varphi_c/T_c>1$ which are obtained using the likelihood function in Eq.~(\ref{lik}).}
  \label{fvctc} 
\end{figure}
\begin{figure}[!bt]
  \centering
  \subfigure{
    \label{fparamdist:a} 
    \includegraphics[width=150pt]{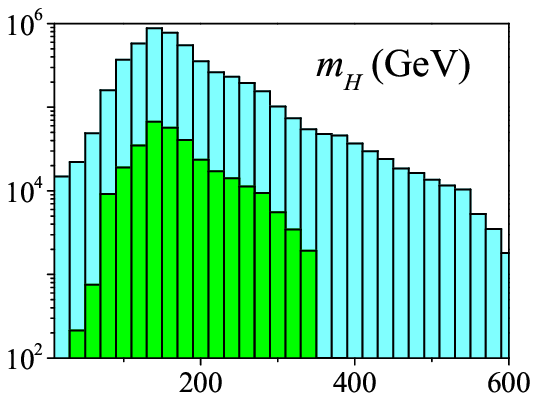}}
  \subfigure{
    \label{fparamdist:b} 
    \includegraphics[width=150pt]{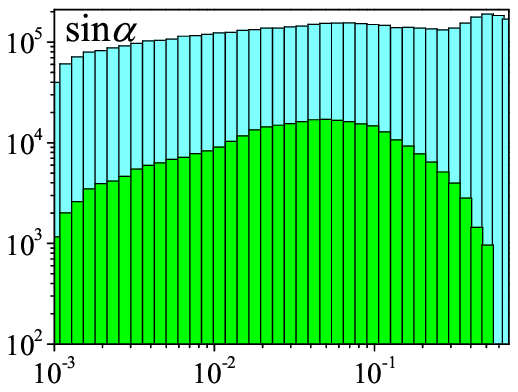}}
  \subfigure{
    \label{fparamdist:c} 
    \includegraphics[width=150pt]{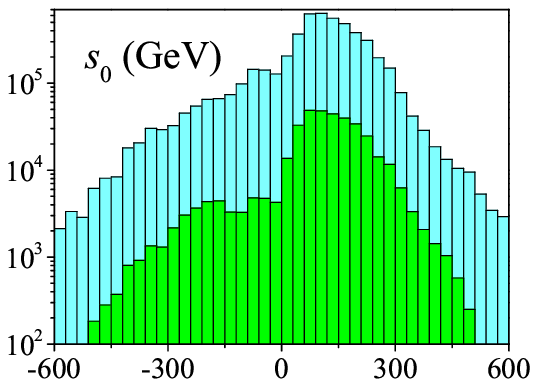}}
  \subfigure{
    \label{fparamdist:d} 
    \includegraphics[width=150pt]{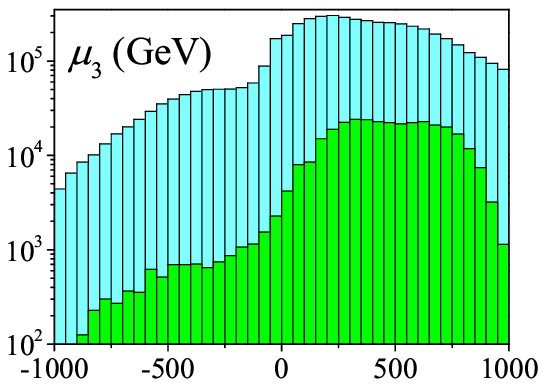}}
  \subfigure{
    \label{fparamdist:e}
    \includegraphics[width=150pt]{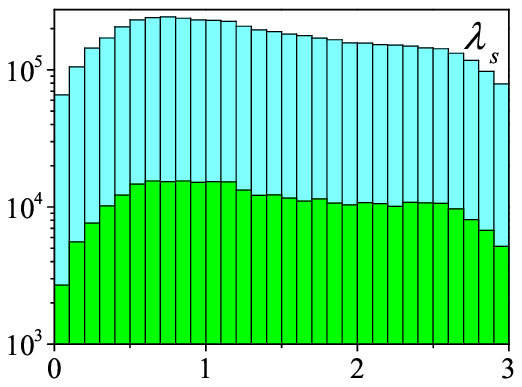}}
  \subfigure{
    \label{fparamdist:f}
    \includegraphics[width=150pt]{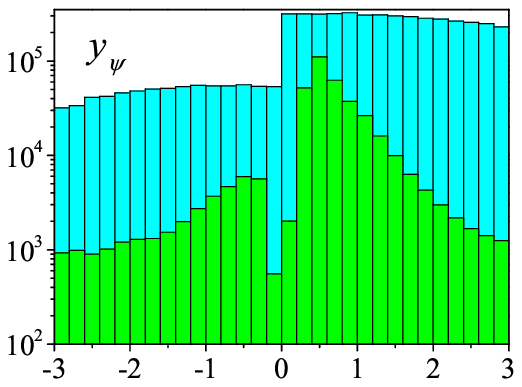}}
  \caption{The frequency distributions of the free parameters $m_H$, $\sin\alpha$, $s_0$, $\mu_3$, $\lambda_s$ and $y_\psi$ of the samples satisfying $\varphi_c/T_c>1$ (cyan areas) which are obtained using the likelihood function in Eq.~(\ref{lik}). The green areas are the distributions after considering all the constraints from the observables such as DM thermal relic density, DM-nucleon cross section, Higgs signal strength, Higg-$Z$-$Z$ coupling strength and the oblique parameters.}
  \label{fparamdist} 
\end{figure}

The relative frequency distribution of the order parameter $\varphi_c/T_c$ is shown in Fig.~\ref{fvctc}. Strongly first order EWPhTs are found with $\varphi_c/T_c$ up to $10$ in this model. The frequency distributions of the $6$ free parameters are shown in Fig.~\ref{fparamdist}. It can be seen that, for $\mathcal{E} = 1$, there exists an upper limit on the mass of the second Higgs particle around $600\GeV$, and $s_0$ is constrained to $\left|s_0\right|\lesssim600\GeV$. Heavier particles cannot trigger a strongly enough first order EWPhT, as the contributions of heavy particles suffer from exponential suppression as shown in Eq.~(\ref{lowT}).

\begin{figure}[!bt]
  \centering
  \subfigure[]{
    \label{fparamspace2:a}
    \includegraphics[width=200pt]{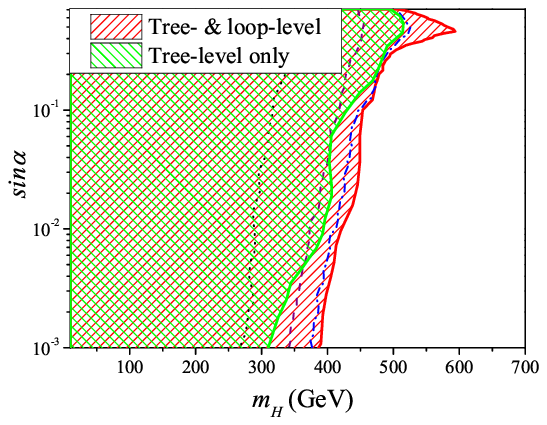}}
  \subfigure[]{
    \label{fparamspace2:c}
    \includegraphics[width=200pt]{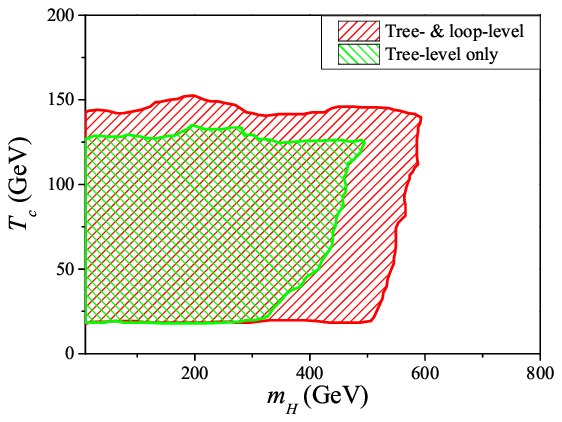}}
  \caption{Comparison between the allowed regions in the $m_H-\sin\alpha$ (left) and $m_H-T_c$ (right) planes for the case with tree-level barrier only and the case with both tree- and loop-level barriers. In the left panel, the lines indicate the boundary of the allowed parameter space for $\varphi_c/T_c > \mathcal{E}$ including both tree- and loop-level barriers with $\mathcal{E} = 1.2$ (blue dot-dashed), $\mathcal{E} = 1.5$ (purple dashed) and $\mathcal{E} = 2$ (black dotted), respectively.}
  \label{fparamspace2}
\end{figure}

In this model, the extra scalar field leads to a tree-level barrier at the critical temperature. Both of the tree- and the loop-level barriers can trigger strongly first order EWPhT. A comparison between the tree- and loop-level barriers is shown in Fig.~\ref{fparamspace2} in which we plot the allowed regions for the case with tree-level barrier only and the case with both tree- and loop-level barriers. As shown by the figure, the allowed region with both the tree- and loop-level barriers is larger than that in tree-level only case. For instance, the upper limits of $m_H$ is about $100\GeV$ higher at $\sin\alpha=0.001$ for $\mathcal{E}=1$. The loop-level cubic terms also raise the critical temperature. As shown in Fig.~\ref{fparamspace2:c}, the critical temperature has an upper limit around $150\GeV$, which is about $10\%$ lower in the case where only the tree-level barrier is considered.

\section{The effect of the sphaleron magnetic moment} \label{sec:sphaleron}

The condition for the sphaleronic interactions to be sufficiently suppressed to preserve the baryon asymmetry generated during the EWPhT is given by~\cite{Shaposhnikov:1987tw,Shaposhnikov:1986jp}
\begin{equation}
\frac{E_{sph}(T_c, B)}{T_c} \gtrsim 35.
\end{equation}
In the SM, the sphaleron energy relates to $\varphi$ the VEV of the Higgs field by
\begin{equation}
E_{\text{sph}} = \frac{8 \pi m_W(\varphi)}{g^2} \mathcal{C},
\end{equation}
where $m_W(\varphi)$ is the $W$-Boson mass, and $\mathcal{C} \sim 2$ is a constant determined by the sphaleron solution. Thus, this condition can be translated into Eq.~(\ref{eq1}) with $\mathcal{E}=1$. This conclusion can be modified if there exists a primordial magnetic field in the early universe~\cite{DeSimone:2011ek,Comelli:1999gt}. The magnetic field can be generated before or during the EWPhT through various mechanisms (for a review see Ref.~\cite{Enqvist:1998fw}). Meanwhile, the electroweak sphaleron may develope a $U(1)_Y$ magnetic dipole moment. The interaction between the magnetic dipole and the background magnetic field can give negative contribution to the sphaleron energy. Consequently, the preservation of the baryon asymmetry requires a larger value of $\mathcal{E}$.

In the presence of a background hypermagnetic field $B$, the sphaleron magnetic moment $\mu$ can lower the sphaleron energy
\begin{equation}
E_{\text{sph}} \left(T, B\right) = E_{\text{sph}}(T) - \mu(T) B.
\end{equation}
In this work, we parametrize the external hypermagnetic field as~\cite{DeSimone:2011ek}
\begin{equation}
B = b T^2,
\end{equation}
where $b$ is a dimensionless parameter which is usually taken to be $b \lesssim 0.4$. To estimate the effect of sphaleron dipole moment in this model, we follow the approach adopted in Refs.~\cite{DeSimone:2011ek,Comelli:1999gt}. The formulas which give the sphaleron solution and the sphaleron magnetic moment are summarized in the Appendix~\ref{sphaleron}.

In this model, the relation between $E_{\text{sph}}/(35T_c)$ and $\varphi_c/T_c$ is complicated and can only be calculated numerically. In Table~\ref{sphaleron_table} we show the values of $E_{\text{sph}}/(35T_c)$ and $\varphi_c/T_c$  for several typical parameter sets. It can be seen that the presence of the sphaleron magnetic moment can lower the sphaleron energy, which makes the value of $E_{\text{sph}}/(35T_c)$ lower than the value of $\varphi_c/T_c$. However, the difference between them are within $10\%$. As can be seen in Table~\ref{sphaleron_table}, in the listed parameter sets, the values of $\varphi_c/T_c$ varies from 1.2 to 2.12, and all of the parameter sets can provide a strongly enough first order EWPhT. This is different from the conclusion in the case of the SM where the inclusion of the magnetic moment generally requires $\varphi_c/T_c \gtrsim 1.3$~\cite{Comelli:1999gt}. The reason is that the extra scalar field $S$ in this model raises the sphaleron energy but gives no contribution to the sphaleron magnetic moment, which weakens the contribution from the sphaleron magnetic moment. In Fig.~\ref{fparamspace2:a} and Fig.~\ref{smallangle}, we show the boundary of the allowed parameter space for $\mathcal{E}=1.2$. It can be seen in Fig.~\ref{smallangle} that, after considering all the constrains from observables, the difference between the upper bound on the mass of the second Higgs particle for $\mathcal{E}=1.2$ and that for $\mathcal{E}=1$ is within $10\hbox{ GeV}$.

\begin{table}
\begin{center}
\begin{tabular}{ccccccccc}
  $m_2$ & $\sin\alpha$ & $s_0$ & $\mu_3$ & $\lambda_s$ & $E_{\text{sph}}(T_c)$ & $E_{\text{dipole}}(T_c, B)$ & $\frac{\varphi_c}{T_c}$ & $\frac{E_{\text{sph}}(T_c, B)}{35T_c}$ \\
  \hline
  256.8 & 0.05 & 42.7 & 125.8 & 1.05 & 1.16 & 0.08 & 1.20 & 1.14 \\
  120.0 & 0.074 & 136.6 & 267.6 & 0.75 & 1.39 & 0.05 & 2.12 & 2.08 \\
  97.2 & 0.002 & 212.8 & 235.5 & 0.70 & 0.90 & 0.06 & 1.21 & 1.16 \\
  197.1 & 0.14 & 100.2 & 464.1 & 0.92 & 1.33 & 0.04 & 1.65 & 1.62 \\
  127.2 & 0.02 & 118.8 & 61.8 & 0.80 & 1.18 & 0.05 & 1.38 & 1.34 \\
\end{tabular}
\end{center}
  \caption{Sphaleron and magnetic dipole energies for several typical parameter sets. The sphaleron energy and magnetic dipole energy are in units of $4\pi\sqrt{\varphi_c^2+s_c^2}/g$. Other parameters $m_2$, $s_0$ and $u_3$ are in unit of GeV. The SM Higgs mass is set to $m_1 = 125 \hbox{ GeV}$. The magnetic field is fixed at $B = 0.4 T^2$. \label{sphaleron_table}}
\end{table}

\section{DM thermal relic density} \label{srelic}

The fermionic DM particle $\psi$ can annihilate into final states $\bar{f}f$, $W^+W^-$, $ZZ$, $hh$, $HH$ or $hH$ via $s$-channel Higgs particle exchanges. For annihilation with final states $hh$, $HH$ or $hH$, the $t$- and $u$-channels are also possible. The Feynman diagrams for these processes are shown in Fig.~\ref{fd_anni}. The cross sections for these processes are given in Appendix~\ref{appB}.

\begin{figure}[!hbt]
  \centering
   \includegraphics[width=420pt]{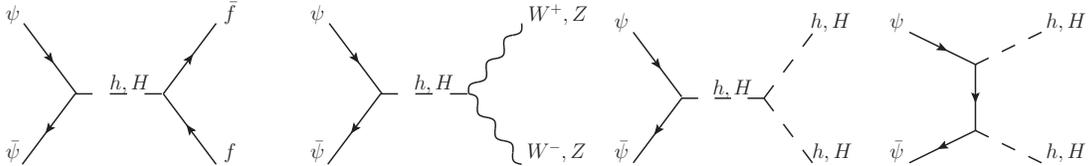}
  \caption{Feynman diagrams for the annihilation of fermionic DM particle.}
   \label{fd_anni}
\end{figure}

The thermal average of the cross section multiplied by the DM relative velocity $v_{\text{rel}}$ at a temperature $T$ is given by
\begin{equation}
\left< \sigma v_{\text{rel}} \right> = \frac{1}{8m_{\psi}^4 T K_2^2 \left(m_{\psi}/T\right)} \int_{4m_{\psi}^2}^{\infty} d\mathfrak{s} \, \sigma \left(\mathfrak{s}\right) \left(\mathfrak{s}-4m_{\psi}^2\right) \sqrt{\mathfrak{s}} \, K_1\left(\frac{\sqrt{\mathfrak{s}}}{T}\right),
\end{equation}
where $K_{1}$ ($K_2$) is the modified Bessel function of the first (second) kind, $\sqrt{\mathfrak{s}}$ denotes the center-of-mass energy. The temperature evolution of the abundance $Y$ which is defined as the number density devided by the entropy density of the DM particle is governed by the Boltzmann equation \cite{Gondolo:1990dk}
\begin{equation} \label{be}
\frac{dY}{dT} = \sqrt{\frac{\pi g_{*}\left(T\right)}{45}} M_{\text{pl}} \langle\sigma v_{\text{rel}}\rangle \left[Y\left(T\right)^2 - Y_{\text{eq}}\left(T\right)^2 \right],
\end{equation}
where $M_{\text{pl}}=1.22 \times 10^{19} \GeV$ is the Planck mass scale, $g_{*}$1 is the effective number of relativistic degrees of freedom, and $Y_{\text{eq}}$ is the abundance at equilibrium. The relic density is related to the present-day abundance $Y\left(T_0\right)$ by
\begin{equation}
\Omega \, h^2 = 2.472 \times 10^8 \GeV^{-1} m_\psi Y\left(T_0\right),
\end{equation}
where $T_0$ is the temperature of the microwave background. In this work we adopt the freeze-out approximation, and use micrOMEGAs3.3 for numerical calculation of the relic density \cite{Belanger:2001fz,Belanger:2006qa}. The freeze-out temperature $T_f$ can be defined from the relation $Y\left(T_f\right) = \left(1+\delta\right)Y_{\text{eq}} \left(T_f\right)$ with $\delta$ being a constant and can be determined by solving
\begin{equation}
\left.\frac{d \ln Y_{\text{eq}}}{d T}\right|_{T=T_f} = \delta\left(\delta+2\right) \sqrt{\frac{\pi g_{*}\left(T_f\right)}{45}} M_{\text{pl}} \langle \sigma v_{\text{rel}} \rangle Y_{\text{eq}}\left(T_f\right) ,
\end{equation}
with $\delta = 1.5$ \cite{Belanger:2001fz}. Below the freeze-out temperature, $Y_{\text{eq}} \ll Y$, Eq.~(\ref{be}) can be integrated
\begin{equation}
\frac{1}{Y\left(T_0\right)} = \frac{1}{Y\left(T_f\right)} + \sqrt{\frac{\pi}{45}} M_{\text{pl}} \int_{T_0}^{T_f} \sqrt{g_{*}\left(T\right)} \langle\sigma v_{\text{rel}}\rangle dT.
\end{equation}
The deviation of this approximation from the exact solution of the Boltzmann equation Eq.~(\ref{be}) is within $2\%$ \cite{Belanger:2001fz}.

\begin{figure}[!tb]
  \centering
  \includegraphics[width=350pt]{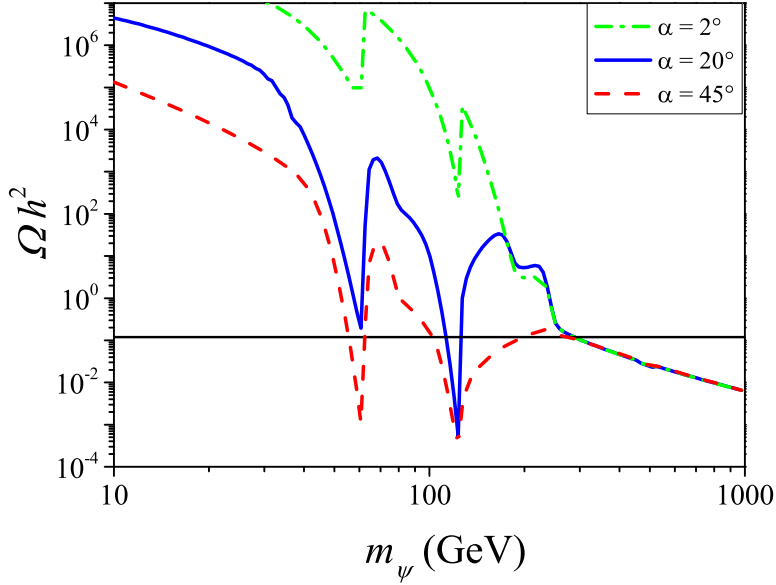}
  \caption{DM thermal relic density as a function of the DM particle mass with $m_H=250\GeV$ for different values of $\alpha=2^\circ$, $20^\circ$ and $45^\circ$, respectively. Other parameters are fixed at $s_0=300\GeV$, $\mu_3=300\GeV$ and $\lambda_s=1$. The horizontal solid line indicates $\Omega h^2=0.1187$ \cite{Ade:2013zuv}.}
   \label{fRelicDensity}
\end{figure}

Fig.~\ref{fRelicDensity} shows the thermal relic density as a function of the DM particle mass. Since the measurement on the DM relic density from the Planck experiment is very precise, the value of $m_\psi$ can actually be solved from the DM relic density up to a five-fold ambiguity. The ambiguity arises from the two resonant annihilations when $m_\psi \approx m_{h,H}/2$.

\section{Direct detection of DM} \label{sdire}

For a Dirac DM particle the spin-independent DM-proton elastic scattering cross section is given by
\begin{equation}
\sigma_{\text{SI}} \approx \frac{m_{r}^{2}}{\pi}\lambda_{p}^{2},
\end{equation}
where $m_r$ is the DM-proton reduced mass $m_r = m_{\psi} m_p / \left(m_{\psi} + m_p\right)$ with $m_p$ the proton mass. The coupling $\lambda_p$ is given by
\begin{equation}
\frac{\lambda_{p}}{m_{p}} = \sum_{q=u,d,s}f_{T_q}^{(p)}\frac{\lambda_{q}}{m_{q}} +\frac{2}{27}f_{T_{g}}^{(p)}\sum_{q=c,b,t}\frac{\lambda_{q}}{m_{q}}.
\end{equation}
The coupling $\lambda_q$ at quark level in this model is
\begin{equation} \label{ddcs}
\frac{\lambda_{q}}{m_{q}} = \frac{y_{\psi}\sin\alpha\cos\alpha}{\varphi_0}\left(\frac{1}{m_h^{2}}-\frac{1}{m_H^{2}}\right).
\end{equation}
The parameters $f_{T_q}^{(p)}$ are defined from the nucleon matrix elements
$m_p\, f_{T_q}^{(p)} \equiv \left<p\left|m_q \bar{q} q\right|p\right>$
for $q = u,d,s$ and $f_{T_{g}}^{(p)} = 1 - \sum_{q=u,d,s}f_{T_{q}}^{(p)}$. In numerical calculations we take the values
$f_{T_{u}}^{(p)} = 0.0153$, $f_{T_{d}}^{(p)} = 0.0191$ and $f_{T_{s}}^{(p)} = 0.0447$ \cite{Belanger:2013oya}. For some of the recent studies of these parameters we refer to the Refs.~\cite{Cheng:2012qr,Crivellin:2013ipa,Cirigliano:2013zta}.

Currently the strongest upper limits on $\sigma_{\text{SI}}$ are given by the LUX experiment \cite{Akerib:2013tjd}. The allowed region in the $m_H-\sin\alpha$ plane is shown in Fig.~\ref{smallangle}. It can be seen that the mixing angle is severely constrained by the LUX data, for instance $\sin\alpha\lesssim0.1$ leading to $\alpha \lesssim 5.7^\circ$ at $m_H=350\GeV$. In the region where $\left|m_H-m_h\right| \lesssim 20\GeV$, the constraint from LUX data is significantly relaxed due to the destructive interference between the contributions from the two Higgs particles, as shown in Eq.~(\ref{ddcs}). In Fig.~\ref{smallangle} we also show the upper bound on the mixing angle corresponding to the data of the XENON100 experiment. It can be seen that the XENON100 constraint on the mixing angle is much weaker than the LUX constraint, for instance $\sin\alpha \lesssim 0.4$ leading to $\alpha\lesssim23^\circ$ at $m_H=350\GeV$.

The next generation of DM direct detection experiments can push the upper bound on $\sigma_{\text{SI}}$ down to $\sim 10^{-47} \text{cm}^2$ \cite{Aprile:2012zx}. This upper bound can further constrain the mixing angle $\alpha$. In Fig.~\ref{smallangle}, we show the upper bound on the mixing angle which corresponds to the projected exclusion limit of the future XENON1T experiment. It can be seen that $\sin\alpha$ can be further constrained to one order of magnitude lower than the upper bound from the LUX data in the regions off resonance, for instance $\sin\alpha \lesssim 0.01$ leading to $\alpha\lesssim0.57^\circ$ at $m_H=350\GeV$.

\section{Higgs signal strength at the LHC} \label{sdiph}

The LHC experiment has reported the discovery of a SM-like Higgs boson \cite{Aad:2012tfa,Chatrchyan:2012ufa}. Throughout our work we take the SM-like Higgs particle mass fixed at $m_h=125\GeV$. The Higgs signal strengths in different channels such as $\bar{b}b$, $\tau^+ \tau^-$, $\gamma \gamma$, $WW^*$ and $ZZ^*$ have been measured by the ATLAS, CMS and CDF experiments. The combined result on the Higgs signal strength with respect to the SM value shows no significant deviation from the SM prediction \cite{Ellis:2013lra}
\begin{equation} \label{eq:HiggsStrength}
r_h = 1.02 ^{+0.11} _{-0.12},
\end{equation}
with $r_h$ defined as the signal strength of the SM-like Higgs particle in new physics models relative to that in the SM. We consider $r_h$ in the range $0.78-1.24$ which corresponds to the approximately $95\%$ confidence level (CL) allowed range.

The signal strength of the SM-like Higgs particle in this model with respect to the SM value is given by
\begin{equation}
r_h = \frac{\sigma_{gg\rightarrow h} B_{h\rightarrow XX}}{\sigma_{gg\rightarrow h}^{\text{SM}} B^{\text{SM}}_{h\rightarrow XX}} = \frac{\sigma_{gg\rightarrow h}}{\sigma_{gg\rightarrow h}^{\text{SM}}} \times \frac{\Gamma_{h \rightarrow XX}}{\Gamma^{\text{SM}}_{h \rightarrow XX}} \times \frac{\Gamma^{\text{SM}}_{h}}{\Gamma_{h}},
\end{equation}
where $X$ stands for any final state particle, $\sigma_{gg\rightarrow h}$ is the production cross section through gluon-gluon fusion of the SM-like Higgs particle, $\Gamma_{h \rightarrow XX}$ is the width of the SM-like Higgs particle decaying to $X$, $\Gamma_h$ is the total decay width of the SM-like Higgs particle, and $\sigma^{\text{SM}}_{gg\rightarrow h}$, $\Gamma^{\text{SM}}_{h \rightarrow XX}$ and $\Gamma_h^{\text{SM}}$ are the corresponding values in the SM.

The mixing between the two Higgs particles leads to a universal $\cos\alpha$ suppression of all the couplings between the SM-like Higgs particle and the SM fermions and gauge bosons, which leads to
\begin{equation}
\frac{\sigma_{gg\rightarrow h}}{\sigma_{gg\rightarrow h}^{\text{SM}}} = \frac{\Gamma_{h\rightarrow XX}}{\Gamma_{h\rightarrow XX}^{\text{SM}}} = \cos^2 \alpha.
\end{equation}
Additionally, the signal strength of the SM-like Higgs particle is also suppressed by two possible new invisible decay channels which are $h \rightarrow \bar{\psi}\psi$ and $h \rightarrow H H$. The total decay width of the SM-like Higgs particle in this model can be written as
\begin{equation}
\Gamma_{h} = \Gamma_{h}^{\text{SM}} \cos^2\alpha + \Gamma_{h\rightarrow \bar{\psi}\psi} + \Gamma_{h \rightarrow H H},
\end{equation}
where $\Gamma_{h\rightarrow \bar{\psi}\psi}$ and $\Gamma_{h \rightarrow H H}$ are the decay widths of the SM-like Higgs particle via the two new channels
\begin{eqnarray}
  \Gamma_{h \rightarrow \bar{\psi}\psi} &=& \frac{y_{\psi}^2 m_h}{8 \pi} \beta_{\psi}^3 \cdot \sin^2\alpha, \\
  \Gamma_{h \rightarrow H H} &=& \frac{\lambda_{hHH}^2}{8\pi m_h} \beta_{H},
\end{eqnarray}
where $\beta_{\psi(H)} = \sqrt{1-4m_{\psi(H)}^2/m_h^2}$ and $\lambda_{hHH}$ is the coupling of $h H H$ defined in Eq.~(\ref{lamhHH}) in Appendix~\ref{appB}. Thus, the signal strength of the SM-like Higgs particle can be written as
\begin{equation} \label{r1}
r_h = \frac{\Gamma^{\text{SM}}_h \, \cos^4\alpha}{\Gamma^{\text{SM}}_h\cos^2\alpha + \Gamma_{h\rightarrow\bar{\psi}\psi} + \Gamma_{h\rightarrow H H}}.
\end{equation}
Note that the signal strength $r_{h}$ is suppressed by $\cos^2{\alpha}$ even if the two new invisible decay channels are kinematically forbidden.

In the parameter region where $m_H < m_h/2$, $\Gamma_{h\rightarrow HH}$ is still considerably large even if the mixing angle is very small. In the limit without mixing between the two Higgs particles, it is given by
\begin{equation}
\Gamma_{h\rightarrow HH} = \frac{\lambda^2 \varphi_0^2}{16\pi m_h} \beta_H,
\end{equation}
which results in a constraint on the parameter $\lambda$
\begin{equation}
\lambda^2 \lesssim 14.2 \frac{m_h \Gamma_h^{\text{SM}}}{\varphi_0^2 \beta_H}.
\end{equation}
In the parameter region where $m_H\lesssim 30\GeV$, this constraint is strong enough to exclude all the sample points, as shown in Fig.~\ref{smallangle}.

Analogously, the signal strength of the second Higgs particle is given by
\begin{equation} \label{r2}
r_H = \frac{\Gamma^{\text{SM}}_H \, \sin^4\alpha}{\Gamma^{\text{SM}}_H\sin^2\alpha + \Gamma_{H\rightarrow\bar{\psi}\psi} + \Gamma_{H\rightarrow h h}}.
\end{equation}
The signal strength of the second Higgs particle is proportional to $\sin^2\alpha$, which comes from the coupling between the second Higgs particle and the SM fermions and gauge bosons, and it is also suppressed by the decay channels $H\rightarrow\bar{\psi}\psi$ and $H\rightarrow hh$.

The allowed region in the $m_H-\sin\alpha$ plane under this constraint is plotted in Fig.~\ref{smallangle}. It can be seen that the result on the signal strength of the SM-like Higgs particle imposes an upper bound on the mixing angle, due to the suppression factor $\cos^2\alpha$ in the signal strength in Eq.~({\ref{r1}}). When the invisible decay of the SM-like Higgs particle through the channel $h\rightarrow HH$ is kinematically forbidden, i.e. $m_H > m_h/2$, the upper limit on the mixing angle is directly given by $\sin^2 \alpha \lesssim 0.22$, leading to $\alpha \lesssim 28^{\circ}$. When the channel $h\rightarrow HH$ is opened, i.e. $m_H < m_h/2$, the mixing angle is further constrained, for instance $\sin\alpha \lesssim 0.01$ leading to $\alpha\lesssim0.57^\circ$ at $m_H=50\GeV$.

Besides the constraint on the signal strength of the SM-like Higgs particle, the current LHC data also set an upper bound on a Higgs particle with a mass larger than $145\GeV$ \cite{Chatrchyan:2013yoa}, which can be translated into an upper bound on $r_H$ in this model. However, this constraint is much weaker than the constraint on $r_h$ as the invisible decay of the second Higgs particle can be very large.

\section{LEP constraint and the electroweak precision test} \label{slep}

The LEP data impose constraints on the ratio of Higgs-$Z$-$Z$ coupling strength with respect of the SM value $\xi^2_\mathcal{H} = \left(g_{\mathcal{H}ZZ}/g_{\mathcal{H}ZZ}^{SM}\right)^2$ with $\mathcal{H}=h,H$, as shown in Fig. 10(a) in Ref.~\cite{Barate:2003sz}. In this model, the Higgs-$Z$-$Z$ coupling strength is suppressed by the mixing between the two Higgs particles
\begin{equation}
\xi^2_h = \cos^2 \alpha, \qquad \xi^2_H = \sin^2 \alpha.
\end{equation}
The allowed region in the $m_H-\sin\alpha$ plane under the constraint from LEP data at $95\%$ CL is shown in Fig.~\ref{smallangle}. This constraint sets an upper bound on the mixing angle in the region with $m_H < 114\GeV$, which is however much weaker compared with that from the LHC and the LUX experiments, as can be seen in the figure.

The second Higgs particle in this model gives extra contributions to the gauge boson self-energy diagrams compared with the SM case, which can affect the oblique parameters $S$, $T$ and $U$ \cite{Peskin:1990zt,Maksymyk:1993zm}. The shifts of the oblique parameters from the SM values $\Delta X \equiv X - X^{SM}$ are given by \cite{Baek:2011aa,Barger:2007im}
\begin{eqnarray}
\Delta T &=& \frac{3}{16\pi s_W^2} \left[ \cos^2 \alpha \left\{ f_T\left(\frac{m_h^2}{m_W^2} \right) - \frac{1}{c_W^2} f_T\left(\frac{m_h^2}{m_Z^2}\right) \right\} \right. \notag \\
& &+ \sin^2 \alpha \left\{ f_T\left(\frac{m_H^2}{m_W^2}\right) - \frac{1}{c_W^2} f_T\left(\frac{m_H^2}{m_Z^2}\right)\right\} \notag \\
& & \left. - \left\{ f_T\left(\frac{m_h^2}{m_W^2}\right) - \frac{1}{c_W^2} f_T\left(\frac{m_h^2}{m_Z^2}\right) \right\} \right] , \\
\Delta S &=& \frac{1}{2\pi} \left[ \cos^2 \alpha f_S \left(\frac{m_h^2}{m_Z^2}\right) + \sin^2 \alpha f_S \left(\frac{m_H^2}{m_Z^2}\right) - f_S \left(\frac{m_h^2}{m_Z^2} \right) \right] , \\
\Delta U &=& \frac{1}{2\pi} \left[ \cos^2 \alpha f_S \left(\frac{m_h^2}{m_W^2}\right) + \sin^2 \alpha f_S \left(\frac{m_H^2}{m_W^2}\right) - f_S \left(\frac{m_h^2}{m_W^2} \right) \right] - \Delta S ,
\end{eqnarray}
where $m_{W(Z)}$ is the masses of the $W$ ($Z$) gauge boson, $c_W^2=m_W^2/m_Z^2$ and $s_W^2=1-c_W^2$. The functions $f_T(x)$ and $f_S(x)$ are defined as
\begin{eqnarray}
f_T\left(x\right) &=& \frac{x \log x}{x-1} , \\
f_S\left(x\right) &=& \left\{
                               \begin{array}{ll}
                                 \frac{1}{12} \left\{ -2x^2+9x + \left[x^2-(6x-18)/x-1+18\right]x\log x \right. \\
                                 \left. +2\sqrt{\left(x-4\right)} \left(x^2-4x+12\right)  \right. \\
                                 \left. \times \left[ \tanh^{-1}\sqrt{x}/\sqrt{x-4} - \tanh^{-1} (x-2)/\sqrt{(x-4)x} \right] \right\},&\text{for } 0<x<4 \vspace{10pt} \\
                                 \frac{1}{12} \left\{ -2x^2+9x + \left[x^2-6x-18/(x-1)+18\right]x\log x \right. \\
                                 \left. +\sqrt{\left(x-4\right)} \left(x^2-4x+12\right) \log \frac{1}{2} \left(x-\sqrt{(x-4)x}-2\right) \right\},&\text{for } x>4.
                               \end{array}
 \right.
\end{eqnarray}

The constraints from the oblique parameters given in Ref. \cite{Barger:2007im,Baak:2011ze} can be translated into constraints on the mass of the second Higgs particle and the mixing angle. We show the $95\%$ CL allowed region in the $m_H-\sin\alpha$ plane in Fig.~\ref{smallangle}. It can be seen that this constraint is weaker compared with the LHC constraint and the LUX constraint.

\section{Combined Results} \label{sresu}

We combine the constraints from all the above mentioned observables such as the DM relic density, the DM-nucleon scattering cross section, the signal strength of the SM-like Higgs particle, the Higgs-$Z$-$Z$ coupling strength and the oblique parameters on the parameter space satisfying $\varphi_c/T_c > 1$. About $2\times10^5$ sample points surviving all the constraints are obtained. The frequency distributions of the 6 free parameters after considering the phenomenological constraints are shown in Fig.~\ref{fparamdist}.

\begin{figure}[!tb]
  \centering
  \includegraphics[width=400pt]{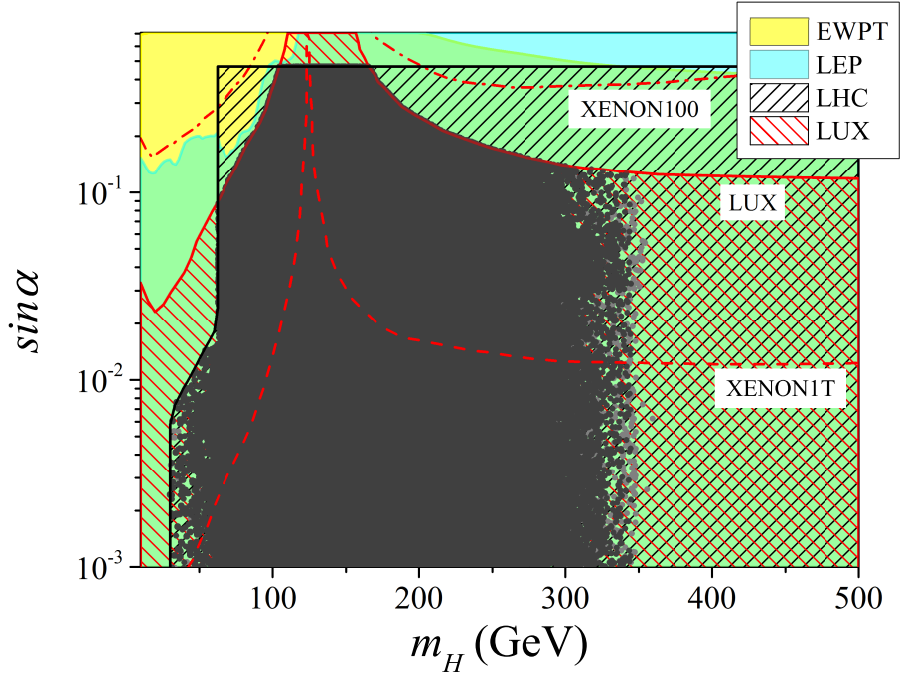}
  \caption{Allowed region in the $m_H-\sin\alpha$ plane satisfying $\varphi_c/T_c>1$ and all the constraints from the electroweak precision test (EWPT) at $95\%$ CL, the LEP data at $95\%$ CL, the Higgs search results at LHC, and the upper bound on DM-nucleon scattering cross section from the LUX experiment. The red dot-dashed line is the upper bound on the mixing angle from the $90\%$ CL XENON100 constraint and the red dashed line is that from the projected exclusion limit of the future XENON1T experiment. The dots are the sample points satisfying $\varphi_c/T_c>\mathcal{E}$ and all the constraints with $\mathcal{E} = 1.2$ (dark gray) and $\mathcal{E} = 1$ (light gray), respectively.}
  \label{smallangle}
\end{figure}

The allowed region in the $m_H-\sin\alpha$ plane is shown in Fig.~\ref{smallangle}. As shown in the figure, the most stringent constraints come from the data of the LHC and the LUX experiments. It can be seen that in the region where the mass of the second Higgs particle is nearly degenerate with that of the SM-like Higgs particle, the LUX constraint is significantly relaxed due to the destructive interference between the contributions from the two Higgs particles. Consequently, in this region the upper bound on the mixing angle is set by the LHC data which leads to $\alpha \lesssim 28^\circ$. In the region where $m_H < m_h/2$, the mixing angle is further constrained, as the invisible decay of the SM-like Higgs particle is opened. In other regions the upper limit on the mixing angle is determined by the LUX data, for instance, $\alpha \lesssim 5.7^\circ$ at $m_H=350\GeV$. As shown by the dots, the requirement of a strongly first order EWPhT sets an upper bound on the mass of the second Higgs particle around $350\GeV$ for $\mathcal{E} = 1$, which is expected as the contributions of very heavy particles to effective potential is suppressed exponentially. As shown by the dark gray dots, when considering $\mathcal{E} = 1.2$ the upper bound on the mass of the second Higgs particle becomes lower. But the difference between the upper bound for $\mathcal{E}=1.2$ and that for $\mathcal{E}=1$ is within $10 \hbox{ GeV}$. A lower bound on the mass of the second Higgs particle around $30\GeV$ is also imposed due to the constraint on $\lambda$ from the LHC data.

The future XENON1T experiment can push the upper bound on $\sigma_{\text{SI}}$ down to $\sim 10^{-47} \text{cm}^2$ \cite{Aprile:2012zx}. The constraint from the projected exclusion limit of the future XENON1T experiment is also shown in Fig.~\ref{smallangle}. It can be seen that a significant proportion of the parameter space can be ruled out by the future XENON1T experiment. The mixing angle can be further constrained to one order of magnitude lower compared with the result of the LUX experiment, for instance $\alpha \lesssim 0.57^\circ$ at $m_H=350\GeV$.

\begin{figure}[!tb]
  \centering
  \includegraphics[width=300pt]{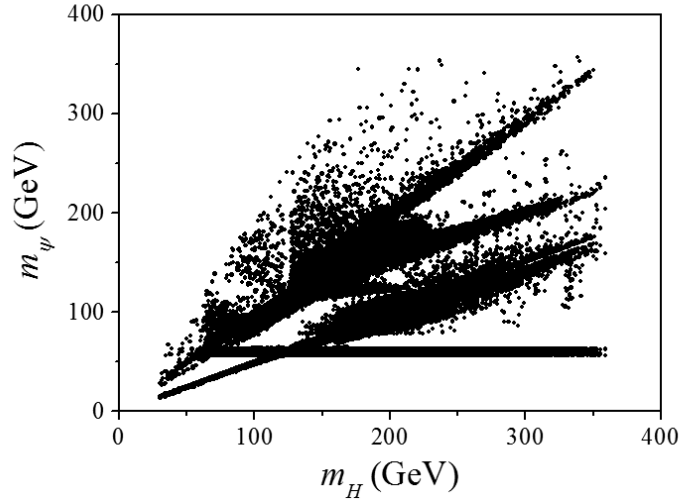}
  \caption{Allowed values of $m_H$ and $m_\psi$ from the sample points satisfying $\varphi_c/T_c>1$ and all the phenomenological constraints (see text for detailed explanation).}
  \label{fmhmp}
\end{figure}

The allowed values of $m_H$ and $m_\psi$ from the sample points are shown in Fig.~\ref{fmhmp}. The DM particle mass is solved from the DM thermal relic density which leads to a five-fold ambiguity. As shown in the figure, there are three branches which correspond to the two resonant annihilations when $m_\psi\approx m_{h,H}/2$ and the threshold of DM annihilation into Higgs particles. It can be seen that the DM particle mass is predicted to be in the range $\sim 15-350\GeV$. The distribution of $y_\psi$ is also significantly changed by the constraint from DM thermal relic density, as shown in Fig.~\ref{fparamdist}.

\section{Conclusion} \label{sconc}

In summary, we have systematically explored the parameter space of the singlet fermionic DM model which can lead to strongly enough first order EWPhT as required by electroweak baryogenesis. We have taken into account the loop-level barrier by including the high temperature approximation up to the terms quartic in $m/T$, and an analytical approximation of the effective potential which well matches both the high- and low-temperature approximations has been introduced, which allows for reliable calculations in low temperature region. It has been shown that the mixing angle is constrained to $\alpha\lesssim 28^\circ$ and the mass of the second Higgs particle is in the range $\sim 30-350\GeV$. The DM particle mass is predicted to be in the range $\sim 15-350\GeV$. The future XENON1T detector can rule out a large proportion of the parameter space. The constraint can be relaxed when the mass of the SM-like Higgs particle is degenerate with that of the second Higgs particle. In other regions the mixing angle can be further constrained to one order of magnitude lower compared with the result using the LUX data, for instance $\alpha \lesssim 0.57^\circ$ at $m_H=350\GeV$.

\section*{Acknowledgement}
This work is supported in part by the National Basic Research Program of China (973 Program) under Grants No.~2010CB833000; the National Nature Science Foundation of China (NSFC) under Grants No.~10975170, No.~10821504, No.~10905084 and No.~11335012; and the Project of Knowledge Innovation Program (PKIP) of the Chinese Academy of Science.

\section*{Appendices}
\appendix
\addcontentsline{toc}{section}{Appendices}
\markboth{APPENDICES}{}
\section{Renormalization of the Higgs potential} \label{appA}
The counter-terms to renormalize the potential at zero temperature are given by
\begin{equation} \label{vct}
V_{\text{CT}}\left(\varphi,s\right) = -\frac{\delta \mu _{\phi}^2}{2} \varphi^2 + \frac{\delta \lambda _{\phi}}{4} \varphi^4 -\frac{\delta \mu _s^2}{2} s^2-\frac{\delta \mu _3}{3} s^3+ \frac{\delta \lambda _s}{4} s^4+\frac{\delta \mu}{2} \varphi^2 s + \frac{\delta \lambda}{4} \varphi^2 s^2 .
\end{equation}
We use the following renormalization conditions
\begin{equation} \label{renorm1}
\left. \left(\frac{\partial}{\partial \varphi}, \frac{\partial}{\partial s}, \frac{\partial^2}{\partial \varphi^2}, \frac{\partial^2}{\partial s^2}, \frac{\partial^2}{\partial s \partial \varphi} \right) \left(V_1\left(\varphi,s\right)+V_{\text{CT}}\left(\varphi,s\right)\right) \right|_{(\varphi,s)=(\varphi_0,s_0)} = 0 ,
\end{equation}
and
\begin{equation} \label{renorm2}
\left(\left.\frac{\partial }{\partial s} \right|_{(\varphi,s)=(0,s_{\varphi})}, \left. \frac{\partial}{\partial v} \right|_{(\varphi,s)=(\varphi_{s},0)} \right) \left(V_1\left(\varphi,s\right)+V_{\text{CT}}\left(\varphi,s\right)\right) = 0,
\end{equation}
where $s_{\varphi}$ ($\varphi_{s}$) is the location of the minimum on the $s$ ($\varphi$) directions. The conditions in Eq.~(\ref{renorm1}) keep the locations of tree-level VEVs and the mass of the two Higgs particles unchanged, and that in Eq.~(\ref{renorm2}) keep the locations of the minima on the $s$ and $\varphi$ direction unchanged. The solutions of the renormalization conditions Eq.~(\ref{renorm1}) and (\ref{renorm2}) are
\begin{eqnarray}
\delta\mu_{\phi}^2 &=& \frac{\varphi_s^3 V_1^{(1,0)}\left(\varphi_0,s_0\right) + 2\varphi_0^3 V_1^{(1,0)}\left(\varphi_s,0\right) - \varphi_0 \varphi_s^3 V_1^{(2,0)}\left(\varphi_0,s_0\right)}{2 \varphi_0^3 \varphi_s} ,\\
\delta\lambda_{\phi} &=& \frac{V_1^{(1,0)}\left(\varphi_0,s_0\right)-\varphi_0 V_1^{(2,0)}\left(\varphi_0,s_0\right)}{2\varphi_0^3} , \\
\delta\mu_s^2 &=& \frac{1}{2s_0^2\left(s_0-s_\varphi\right)^2s_\varphi} \left\{s_0 \left[2s_0^3 V_1^{(0,1)}\left(0,s_\varphi\right)+s_\varphi^2\left(-6s_0+4s_\varphi\right)V_1^{(0,1)}\left(\varphi_0,s_0\right) \right. \right. \notag \\
& &\left.+ 2s_\varphi^2s_0\left(s_0-s_\varphi\right)V_1^{(0,2)}\left(\varphi_0,s_0\right) + s_\varphi^2 \varphi_0\left(2s_0-s_\varphi\right)V_1^{(1,1)}\left(\varphi_0,s_0\right)\right]  \notag \\
& & \left. - \varphi_0^2 s_0^2 s_\varphi^2 \left(s_0-s_\varphi\right) \delta\lambda \right\} ,
\end{eqnarray}
\begin{eqnarray}
\delta\mu_3 &=& \frac{1}{2s_0^3\left(s_0-s_\varphi\right)^2 s_\varphi} \left\{ -2s_0 \left[2s_0^3 V_1^{(0,1)}\left(0,s_\varphi\right)+s_\varphi\left(-3s_0^2+s_\varphi^2\right)V_1^{(0,1)}\left(\varphi_0,s_0\right) \right. \right. \notag \\
& &\left.+ s_\varphi s_0\left(s_0^2-s_\varphi^2\right)V_1^{(0,2)}\left(\varphi_0,s_0\right) + s_\varphi \varphi_0 s_0^2 V_1^{(1,1)}\left(\varphi_0,s_0\right)\right]  \notag \\
& & \left. - \varphi_0^2 s_0^2 s_\varphi \left(s_0^2-s_\varphi^2\right) \delta\lambda \right\} ,
\end{eqnarray}
\begin{eqnarray}
\delta\lambda_s &=& \frac{1}{2s_0^3\left(s_0-s_\varphi\right)^2 s_\varphi} \left\{ -s_0 \left[2s_0^2 V_1^{(0,1)}\left(0,s_\varphi\right)+s_\varphi\left(-4s_0+2s_\varphi\right)V_1^{(0,1)}\left(\varphi_0,s_0\right) \right. \right. \notag \\
& &\left.+ 2s_\varphi s_0\left(s_0-s_\varphi\right)V_1^{(0,2)}\left(\varphi_0,s_0\right) + s_\varphi \varphi_0 s_0 V_1^{(1,1)}\left(\varphi_0,s_0\right)\right]  \notag \\
& & \left. - \varphi_0^2 s_0^2 s_\varphi \left(s_0-s_\varphi\right) \delta\lambda \right\} ,\\
\delta\mu &=& -\delta\lambda s_0  ,\\
\delta\lambda &=& \frac{1}{\varphi_0^3 s_0 \varphi_s} \left[\left(3\varphi_0^2 \varphi_s-\varphi_s^3\right)V_1^{(1,0)}\left(\varphi_0,s_0\right) - 2\varphi_0^3 V_1^{(1,0)}\left(\varphi_s,0\right) \right. \notag \\
& & \left.- \varphi_0^2 s_0 \varphi_s V_1^{(1,1)}\left(\varphi_0,s_0\right) - \varphi_0 \varphi_s\left(\varphi_0^2-\varphi_s^2\right)V_1^{(2,0)}\left(\varphi_0,s_0\right)\right],
\end{eqnarray}
where
\begin{equation}
V^{(m,n)}\left(\varphi,s\right) = \frac{\partial^{(m+n)}V\left(\varphi,s\right)}{\partial h^m \partial s^m}.
\end{equation}

\section{Cross sections for DM annihilation} \label{appB}

The cross sections for DM particles annihilating into the SM fermions and gauge bosons are given by \cite{Kim:2008pp}
\begin{multline} \label{relic}
\sigma v_{\text{rel}}\left(\bar{\psi}\psi\rightarrow \bar{f}f,W^+W^-,ZZ\right) =  \frac{\left(y_\psi \sin\alpha \cos\alpha\right)^2}{16\pi} \left(1 - \frac{4m_{\psi}^2}{\mathfrak{s}}\right) \\
\times \left| \frac{1}{\mathfrak{s}-m_{h}^2 + i m_{h}\Gamma_{h}} + \frac{1}{\mathfrak{s}-m_{H}^2 + i m_{H}\Gamma_{H}}\right|^2 \cdot A_{f,W,Z},
\end{multline}
where $\Gamma_h$ ($\Gamma_H$) is the total decay width of the SM-like Higgs particle (the second Higgs particle), $\sqrt{\mathfrak{s}}$ denotes the center-of-mass energy, and $A_{f,W,Z}$ stands for the contributions from channels with final states $\bar{f}f$, $W^+W^-$ and $ZZ$
\begin{eqnarray}
A_f &=& 6\, \mathfrak{s} \left(\frac{m_f}{\varphi_0}\right)^2 \times \left(1-\frac{4m_f^2}{\mathfrak{s}}\right)^{3/2}, \\
A_{W} &=& 4 \, \left(\frac{m_{W}^2}{\varphi_0}\right)^2 \times \left(2+\frac{(\mathfrak{s}-2m_{W}^2)^2}{4m_{W}^4}\right) \times \sqrt{1-\frac{4m_{W,Z}^2}{\mathfrak{s}}} .
\end{eqnarray}
$A_{Z}$ is defined analogously with $A_{W}$ and there is an additional factor of $1/2$ for $A_Z$.

The cross sections for DM particles annihilating into two identical Higgs particles through $s$-channele are given by \cite{Qin:2011za}
\begin{equation}
\sigma v_{\text{rel}}^{(\text{s})}\left(\bar{\psi}\psi\rightarrow\mathcal{H}\mathcal{H}\right) = \frac{1}{2} \kappa_{\mathcal{H}} \left(\mathfrak{s} - 4m_\psi^2\right) \left| \frac{y_{h} \lambda_{h\mathcal{H}\mathcal{H}}}{\mathfrak{s}-m_{h}^2 + i m_{h}\Gamma_{h}} + \frac{y_{H} \lambda_{H\mathcal{H}\mathcal{H}}}{\mathfrak{s}-m_{H}^2 + i m_{H}\Gamma_{H}}\right|^2,
\end{equation}
where $\mathcal{H}$ stands for $H$ or $h$, and $\kappa_{\mathcal{H}}$ is defined as
\begin{equation}
\kappa_{\mathcal{H}} = \frac{1}{16\pi\,\mathfrak{s}^2} \sqrt{\mathfrak{s}^2 -4\mathfrak{s} m_{\mathcal{H}}^2},
\end{equation}
The cross sections for DM particles annihilating into two identical Higgs particles through $t$- and $u$-channel are given by
\begin{eqnarray}
\sigma v_{\text{rel}}^{(\text{t+u})}\left(\bar{\psi}\psi\rightarrow\mathcal{H}\mathcal{H}\right) &=& \kappa_{\mathcal{H}} \, y_{\mathcal{H}}^4 \left\{ \frac{\left(4 m_\psi^2 - m_{\mathcal{H}}^2\right)^2}{D^2-A^2} - \log\left|\frac{A+D}{A-D}\right| \left[ \frac{\left(\mathfrak{s}+8m_\psi^2-2m_{\mathcal{H}}^2\right)}{2D} \right. \right. \nonumber \\
 & & \left. \left.+\frac{\left(16m_\psi^4 -4m_\psi^2 \mathfrak{s}-m_{\mathcal{H}}^4\right)}{AD}\right]-2 \right\},
\end{eqnarray}
where $A$ and $D$ are defined as
\begin{equation}
A = \frac{1}{2}\left(2 m_{\mathcal{H}}^2 - \mathfrak{s}\right), \qquad D=\frac{\mathfrak{s}}{2}\beta_\psi \beta_{\mathcal{H}},
\end{equation}
with $\beta_{\psi} = \sqrt{1-4m_\psi^2 / \mathfrak{s}}$ and $\beta_{\mathcal{H}} = \sqrt{1-4m_{\mathcal{H}}^2/\mathfrak{s}}$.
The interference terms between the $s$- and $u$-, $t$-channels are given by
\begin{eqnarray}
\sigma v_{\text{rel}}^{(\text{int})}\left(\bar{\psi}\psi\rightarrow\mathcal{H}\mathcal{H}\right) &=& 2\kappa_{\mathcal{H}}\, y_{\mathcal{H}}^2\, m_{\psi} \left[ \frac{y_{h} \lambda_{h\mathcal{H}\mathcal{H}} \left(\mathfrak{s}-m_{h}^2\right)}{ \left(\mathfrak{s}-m_{h}^2\right)^2 + m_{h}^2 \Gamma_{h}^2} + \frac{y_{H} \lambda_{H\mathcal{H}\mathcal{H}} \left(\mathfrak{s}-m_{H}^2\right)}{ \left(\mathfrak{s}-m_{H}^2\right)^2 + m_{H}^2 \Gamma_{H}^2}\right] \nonumber \\
 & & \times \log\left|\frac{A+D}{A-D}\right|\left( \frac{A}{D}+\frac{1}{2}\frac{\beta_\psi}{\beta_{\mathcal{H}}}-2\right).
\end{eqnarray}

The cross sections for DM particles annihilating into $h$ and $H$ through $s$-channel are given by
\begin{equation}
\sigma v_{\text{rel}}^{(\text{s})}\left(\bar{\psi}\psi\rightarrow hH\right) = \kappa_{hH} \left(\mathfrak{s} - 4m_\psi^2\right) \left| \frac{y_{h} \lambda_{Hhh}}{\mathfrak{s}-m_{h}^2 + i m_{h}\Gamma_{h}} + \frac{y_{H} \lambda_{hHH}}{\mathfrak{s}-m_{H}^2 + i m_{H}\Gamma_{H}}\right|^2,
\end{equation}
where $\kappa_{hH}$ is defined as
\begin{equation}
\kappa_{hH} = \frac{1}{16\pi\,\mathfrak{s}^2} \sqrt{\mathfrak{s}^2 -2\mathfrak{s}\left(m_{h}^2+m_{H}^2\right) + \left(m_{h}^2-m_{H}^2\right)^2}.
\end{equation}
The cross sections for DM particle annihilation into $h$ and $H$ through $t$- and $u$-channel are given by
\begin{eqnarray}
\sigma v_{\text{rel}}^{(\text{t+u})}\left(\bar{\psi}\psi\rightarrow hH\right) &=& 2 \kappa_{hH}\, y_{h}^2 \, y_{H}^2 \left\{ \frac{\left(4 m_\psi^2 - m_{h}^2\right)\left(4 m_\psi^2 - m_{H}^2\right)}{D^2-A^2} \right. \nonumber \\
 & & \left. - \log\left|\frac{A+D}{A-D}\right| \left[ \frac{\left(\mathfrak{s}+8m_\psi^2-m_h^2-m_H^2\right)}{2D} \right. \right. \nonumber \\
 & & \left. \left.+\frac{\left(16m_\psi^4 -4m_\psi^2 \mathfrak{s}-m_h^2 m_H^2\right)}{AD}\right]-2 \right\},
\end{eqnarray}
where $A$ and $D$ are defined as
\begin{equation}
A = \frac{1}{2}\left( m_h^2 + m_H^2 - \mathfrak{s}\right), \qquad D=\frac{\mathfrak{s}}{2}\beta_\psi \beta_{hH},
\end{equation}
with
\begin{equation*}
\beta_{hH} = \sqrt{1-\frac{\left(m_h+m_H\right)^2}{\mathfrak{s}}}\sqrt{1-\frac{\left(m_h-m_H\right)^2}{\mathfrak{s}}}.
\end{equation*}
The interference terms between the $s$- and $u$-, $t$-channels are given by
\begin{eqnarray}
\sigma v_{\text{rel}}^{(\text{int})}\left(\bar{\psi}\psi\rightarrow hH\right) &=& 4\kappa_{hH}\, y_h\, y_H\, m_{\psi} \left[ \frac{y_{h} \lambda_{Hhh} \left(\mathfrak{s}-m_{h}^2\right)}{ \left(\mathfrak{s}-m_{h}^2\right)^2 + m_{h}^2 \Gamma_{h}^2} + \frac{y_{H} \lambda_{hHH} \left(\mathfrak{s}-m_{H}^2\right)}{ \left(\mathfrak{s}-m_{H}^2\right)^2 + m_{H}^2 \Gamma_{H}^2}\right] \nonumber \\
 & & \times \log\left|\frac{A+D}{A-D}\right|\left( \frac{A}{D}+\frac{1}{2}\frac{\beta_\psi}{\beta_{hH}}-2\right).
\end{eqnarray}

The physical couplings in this model are given by
\begin{eqnarray}
y_{\mathcal{H}} &=& \left\{
              \begin{array}{ll}
                y_\psi\sin\alpha, & \hbox{if $\mathcal{H}=h$;} \\
                y_\psi\cos\alpha, & \hbox{if $\mathcal{H}=H$.}
              \end{array}
 \right. \nonumber \\
\lambda_{hhh} &=& c_\alpha^3 \lambda_\phi \varphi_0 - \frac{1}{2} c_\alpha^2 s_\alpha \lambda s_0 - \frac{1}{2} c_\alpha^2 s_\alpha \mu + \frac{1}{2} c_\alpha s_\alpha^2 \lambda \varphi_0 - s_\alpha^3 \lambda_s s_0 + \frac{1}{3} s_\alpha^3 \mu_3, \nonumber \\
\lambda_{hHH} &=& c^2_{\alpha} s_{\alpha}\lambda s_0-\frac{1}{2}s^3_{\alpha} \lambda s_0+\frac{1}{2}c^3_{\alpha} \lambda \varphi_0- c_{\alpha} s^2_{\alpha} \lambda \varphi_0-3 c^2_{\alpha} s_{\alpha} \lambda_s s_0  \notag \\
& &  +3 c_{\alpha} s^2_{\alpha} \lambda_{\phi} \varphi_0 + c^2_{\alpha} s_{\alpha} \mu-\frac{1}{2}s^3_{\alpha} \mu+ c^2_{\alpha} s_{\alpha} \mu_{3} , \label{lamhHH} \\
\lambda_{Hhh} &=& \frac{1}{2} c_\alpha^3\lambda s_0 + \frac{1}{2} c_\alpha^3 \mu - c_\alpha^2 s_\alpha \lambda \varphi_0 + 3 c_\alpha^2 s_\alpha \lambda_\phi \varphi_0 - c_\alpha s_\alpha^2 \lambda s_0 \nonumber \\
& & + 3 c_\alpha s_\alpha^2 \lambda_s s_0 - c_\alpha s_\alpha^2 \mu - c_\alpha s_\alpha^2 \mu_3 + \frac{1}{2} s_\alpha^3 \lambda \varphi_0, \nonumber \\
\lambda_{HHH} &=& c_\alpha^3 \lambda_s s_0 - \frac{1}{3} c_\alpha^3 \mu_3 + \frac{1}{2} c_\alpha^2 s_\alpha \lambda \varphi_0 + \frac{1}{2} c_\alpha s_\alpha^2 \lambda s_0 + \frac{1}{2} c_\alpha s_\alpha^2 + s_\alpha ^3 \lambda_\phi \varphi_0. \nonumber
\end{eqnarray}
where $c_\alpha$ and $s_\alpha$ stand for $\cos\alpha$ and $\sin\alpha$, respectively.

\section{Sphaleron solution with magnetic moment} \label{sphaleron}

The Lagrangian of the gauge and Higgs sectors of the singlet fermionic DM model is given by
\begin{equation}
\mathcal{L} = - \frac{1}{4} F^a_{\mu\nu} F^{a, \mu\nu} - \frac{1}{4} f_{\mu\nu} f^{\mu\nu} + \left(D_\mu \Phi\right)^{\dagger} \left(D^{\mu}\Phi\right) + \frac{1}{2} \partial_\mu S \partial^{\mu} S - V\left(\Phi, S, T\right),
\end{equation}
where
\begin{eqnarray}
F_{\mu\nu}^a &=& \partial_\mu W_\nu^a - \partial_\nu W_\mu^a + g \epsilon^{abc} W_\mu^b W_\nu^c, \nonumber \\
f_{\mu\nu} &=& \partial_\mu a_\nu - \partial_\nu a_\mu, \nonumber \\
D_\mu &=& = \partial_\mu - \frac{i}{2}g \sigma^a W_\mu^a - \frac{i}{2}g^{\prime} a_\mu H, \nonumber
\end{eqnarray}
where $W^a_\mu (a = 1,2,3)$ and $a_\mu$ are the $SU(2)_L$ and $U(1)_Y$ gauge fields, respectively. The Higgs potential $V\left(\Phi, S, T\right)$ is the effective potential at temperature $T$. The corresponding energy functional is given by
\begin{equation}
E = \int d^3 x \left[ \frac{1}{4} F_{ij}^a F_{ij}^a + \left(D_i\phi\right)^\dagger \left(D_i\phi\right) + \frac{1}{2} \partial_i S \partial_i S + V\left(\Phi, S, T\right) \right].
\end{equation}

In the limit of vanishing weak mixing angle, $\theta_w \approx 0$, the $U(1)_Y$ gauge field decouples, and the sphaleron solution is spherically symmetric. We adopt the ansatz for the fields from Refs.~\cite{Klinkhamer:1984di,Klinkhamer:1990fi,Enqvist:1992kd,Choi:1994mf}
\begin{eqnarray}
g W^a_i \sigma^a d x^i &=& \left(1-f(\xi)\right)F_a \sigma^a, \\
\Phi &=& \frac{\varphi}{\sqrt{2}} \left(
                                    \begin{array}{c}
                                      0 \\
                                      h(\xi) \\
                                    \end{array}
                                  \right), \\
S &=& s \, p(\xi),
\end{eqnarray}
where $\xi \equiv gvr$ is the dimensionless distance, and the functions $F_a$ are defined as~\cite{Klinkhamer:1990fi}
\begin{eqnarray}
F_1 &=& -2 \sin\phi d\theta - \sin 2\theta \cos\phi d\phi, \\
F_2 &=& -2 \cos\phi d\theta + \sin 2\theta \sin\phi d\phi, \\
F_3 &=& 2\sin^2\theta d\phi.
\end{eqnarray}
The sphaleron energy can be minimized by the solving the variational field equations
\begin{eqnarray}
f^{\prime\prime} &=& \frac{2}{\xi^2}f(f-1)(1-2f) + \frac{1}{4} h^2 (f-1), \label{eq:0_sph_f} \\
h^{\prime\prime} + \frac{2}{\xi} h^{\prime} &=& \frac{2}{\xi^2} h (1-f)^2 + \frac{1}{g^2 \varphi^4} \frac{\partial V(h, p, T)}{\partial h}, \label{eq:0_sph_h} \\
p^{\prime\prime} + \frac{2}{\xi} p^{\prime} &=& \frac{1}{g^2 \varphi^2 s^2} \frac{\partial V(h,p,T)}{\partial p}, \label{eq:0_sph_p}
\end{eqnarray}
where the prime denotes the derivative with respect to $\xi$. To ensure the smoothness at the origin and the asymptotic behavior at $\xi \rightarrow \infty$, the boundary conditions for $f(\xi)$, $h(\xi)$ and $p(\xi)$ are given by
\begin{equation}
f(0) = h(0) = 0,
\end{equation}
and
\begin{equation}
f(\infty) = h(\infty) = p(\infty) = 1.
\end{equation}
Note that the value of $S$ at the origin is not constrained by any condition. The boundary condition for $p(\xi)$ can be obtained from the Taylor expansion of the equations around $\xi=0$, which leads to $p^{\prime}(0) = 0$.

For non-vanishing weak mixing angle, $\theta_W\neq 0$, the $U(1)_Y$ gauge field must be taken into account because its source term is nonzero. The source term of the $U(1)_Y$ gauge field $a_i$ is given by the current
\begin{equation}
\partial_{ij} f_{ij} = J_i = - \frac{i}{2} g^{\prime} \left[\Phi^\dagger D_i \Phi - \left(D_i\Phi\right)^\dagger \Phi \right].
\end{equation}
At the leading order in $\theta_W$, $a_i$ in the current can be neglected, which leads to
\begin{equation}
J_i = -\frac{1}{2} g^\prime \varphi^2 \frac{1}{r^2} h^2(\xi) \left[1-f(\xi)\right] \epsilon_{3ij} x_j.
\end{equation}
Thus, in the presence of a constant background magnetic field $B$ along the $z$-axis, the energy of the $U(1)_Y$ field is given by
\begin{equation} \label{edipole}
E = - \int d^3 x a_i^{\text{bg}} J_i,
\end{equation}
where $a_i^{\text{bg}} = -(B/2)\epsilon_{3ij} x_j$ is the vector potential of the background magnetic field. The sphaleron energy in Eq.~(\ref{edipole}) can be rewritten in the form of a magnetic moment $\mu$ along the $z$-axis in the background magnetic field
\begin{equation}
E = E_{\text{dipole}} = - \mu B,
\end{equation}
where the magnetic moment $\mu$ is defined as
\begin{equation}
\mu = \frac{2\pi}{3} \frac{g^{\prime}}{g^3 \varphi(T)} \int_0^\infty d\xi \xi^2 h^2(\xi) [1-f(\xi)].
\end{equation}

Thus, the non-vanishing weak mixing angle gives rise to a sphaleron magnetic moment~\cite{Klinkhamer:1984di}, and the sphaleron solution becomes axially symmetric~\cite{Manton:1983nd}. In this case, the ansatz for the fields can be chosen as~\cite{Klinkhamer:1990fi}
\begin{eqnarray}
g^{\prime} a_i dx^i &=& \left[1-f_0\left(\xi\right)\right] F_3, \\
g W_i^a \sigma^a dx^i &=& \left[1-f\left(\xi\right)\right] \left(F_1\sigma^1 + F_2\sigma^2\right) + \left[1-f_3\left(\xi\right)\right] F_3 \sigma^2, \\
\Phi &=& \frac{\varphi}{\sqrt{2}} \left(
                                    \begin{array}{c}
                                      0 \\
                                      h(\xi) \\
                                    \end{array}
                                  \right),  \\
S &=& s \, p(\xi),
\end{eqnarray}
with $i = 1,2,3$.
The energy functional is
\begin{eqnarray}
E &=& \frac{4 \pi \varphi}{g} \int_0^\infty d\xi \left\{ \frac{8}{3} f^{\prime 2} + \frac{4}{3} f_3^{\prime 2} + \frac{8}{\xi^2} \left[ \frac{2}{3} f_3^2\left(1-f\right)^2 + \frac{1}{3} \left(f(1-f)+f-f_3\right)^2\right] \right. \nonumber \\
& & \left. + \frac{4g^2}{3g^{\prime 2}} \left[f_0^{\prime 2} + \frac{2}{\xi^2} \left(1-f_0\right)^2 \right] + \frac{1}{2} \xi^2 h^{\prime 2} + h^2 \left[\frac{1}{3} \left(f_0-f_3\right)^2 + \frac{2}{3}\left(1-f\right)^2 \right] \right. \nonumber \\
& & \left. + \frac{s^2}{2\varphi^2} \xi^2 p^{\prime 2} + \frac{\xi^2}{g^2\varphi^4} V\left(h, p, T\right)  \right\}.
\end{eqnarray}
The energy functional can be minimized by solving the variational equations
\begin{eqnarray}
f^{\prime\prime} &=& \frac{2}{\xi^2}(f-1)[f(f-2) + f_3(1+f_3)] + \frac{1}{4} h^2 (f-1), \\
f_3^{\prime\prime} &=& \frac{2}{\xi^2}[3f_3+f(f-2)(1+2f_3)]+\frac{1}{4}h^2(f_3-f_0), \\
f_0^{\prime\prime} &=& \frac{2}{\xi^2}(f_0-1) + \frac{g^{\prime 2}}{4 g^2} h^2 (f_0-f_3), \\
h^{\prime\prime} + \frac{2}{\xi} h^{\prime} &=& \frac{2}{3\xi^2} h [2(1-f)^2 + (f_0-f_3)^2] + \frac{1}{g^2 \varphi^4} \frac{\partial V(h, p, T)}{\partial h}, \\
p^{\prime\prime} + \frac{2}{\xi} p^{\prime} &=& \frac{1}{g^2 \varphi^2 s^2} \frac{\partial V(h,p,T)}{\partial p},
\end{eqnarray}
with boundary conditions given by
\begin{equation}
f(0) = f_3(0) = h(0) = 0, \quad f_0(0) = 1, \quad p^{\prime}(0) = 0,
\end{equation}
and
\begin{equation}
f(\infty) = f_3(\infty) = f_0(\infty) = h(\infty) = p(\infty) = 1.
\end{equation}

\bibliographystyle{JHEP}

\providecommand{\href}[2]{#2}
\begingroup\raggedright\endgroup

\end{document}